\documentclass[10pt,journal,compsoc]{IEEEtran}
\newif\ifpeerreview

\peerreviewfalse

\usepackage[square,sort,comma,numbers]{natbib}
\usepackage{url}
\usepackage{amsmath,amssymb,graphicx}

\usepackage{hyperref}
\usepackage{lipsum} 

\usepackage[switch]{lineno}

\usepackage{xcolor,comment}
\definecolor{mygreen}{RGB}{0, 176, 80}
\definecolor{myyellow}{RGB}{255, 192, 0}
\definecolor{myblue}{RGB}{91, 155, 213}

\usepackage{subfiles} 
\usepackage{caption}
\usepackage{subcaption}

\usepackage{enumitem}
\setlist{leftmargin=*}

\newcommand{\paperID}{46}   

\title{PS$^2$F: Polarized Spiral Point Spread Function for Single-Shot 3D Sensing}

\author{Bhargav~Ghanekar, Vishwanath Saragadam, Dushyant Mehra, Anna-Karin Gustavsson, \\Aswin C. Sankaranarayanan, Ashok Veeraraghavan
\IEEEcompsocitemizethanks{\IEEEcompsocthanksitem B.G., V.S., A.V. are with the Department
of Electrical and Computer Engineering, Rice University, Houston,
TX.\protect\\
E-mail: bhargav.ghanekar@rice.edu
\IEEEcompsocthanksitem D.M., A.-K.G. are with the Department of Chemistry, Rice University, Houston TX.
\IEEEcompsocthanksitem A.C.S. is with the Department of Electrical and Computer Engineering, Carnegie Mellon University, Pittsburgh PA.
}

}

\begin{document}

\IEEEtitleabstractindextext{%
\begin{abstract}
%
We propose a compact snapshot monocular depth estimation technique that relies on an engineered point spread function (PSF).
Traditional approaches used in microscopic super-resolution imaging such as the Double-Helix PSF (DHPSF) are ill-suited for scenes that are more complex than a sparse set of point light sources.
We show, using the Cram\'er-Rao lower bound,   that separating the two lobes of the DHPSF and thereby capturing two separate images leads to a dramatic increase in depth accuracy.
A special property of the phase mask used for generating the DHPSF is that a separation of the phase mask into two halves leads to a spatial separation of the two lobes. 
We leverage this property to build a compact polarization-based optical setup, where we place two orthogonal linear polarizers on each half of the DHPSF phase mask and then capture the resulting image with a polarization-sensitive camera.
Results from simulations and a lab prototype demonstrate that our technique achieves up to $50\%$ lower depth error compared to state-of-the-art designs including the DHPSF and the Tetrapod PSF, with little to no loss in spatial resolution.
\end{abstract}

\begin{IEEEkeywords} 
Computational Photography, 3D Sensing, Microscopy, Phase Mask Design, Polarization-encoded PSFs
\end{IEEEkeywords}
}

\maketitle
\thispagestyle{empty}

\IEEEraisesectionheading{
  \section{Introduction}\label{sec:introduction}
}
%
%
%
%


\IEEEPARstart{3}{D} scanning
is crucial to a wide range of applications, including microscopy~\cite{fischer2011microscopy}, autonomous driving~\cite{arnold2019survey}, and robot-assisted surgeries~\cite{reiter2014surgical}. 
%
Among the multitude of approaches to measure depth, perhaps the hardest are those that involve passive, monocular, and snapshot measurements---a scenario that prioritizes compactness and time resolution. Estimating 3D information in this context  relies on  cues such as shading~\cite{ping1994shape} or defocus~\cite{levin2007image, zhou2009coded}; however, the underlying inverse problem is challenging and ill-posed.
A key property of real cameras is that the defocus blur changes with the depth of scene points, which has been leveraged by prior work to obtain depth. 
However, the blur produced by conventional pupils is not  conducive to robust depth estimation; to mitigate this, there has been significant interest in the design of engineered pupil plane masks in the form of
an amplitude ~\cite{levin2007image}, or phase mask~\cite{pavani2009three,greengard2006depth,pavani2008high, shechtman2014optimal,badieirostami2010three,nehme2020deepstorm3d,nehme2020learning}.
In particular, the seminal work of Pavani \textit{et al.} \cite{pavani2008high} has demonstrated that the so-called Double-Helix PSF (DHPSF), consisting of two lobes rotating about a center, can provide high spatial and depth resolutions, at least in the context of super-resolution localization microscopy. 

The majority of prior work on engineered PSFs focus on isolated point light sources; this is a consequence of their intended application---namely super-resolved localization of fluorescent particles---and the simplification that this assumption provides.
%
However, real-world scenes often consist of more complex geometric primitives such as lines, edges, and curves~\cite{cho2011blur,joshi2008psf}; a PSF optimized for  point sources is  inadequate at recovering such complex geometries due to ambiguities in depth estimation.
This is all the more important in applications involving linear structures including blood vasculature, neuronal networks (of the biological kind), and microtubules.

%

\begin{figure*}[!t]
\centering
\includegraphics[width=\textwidth]{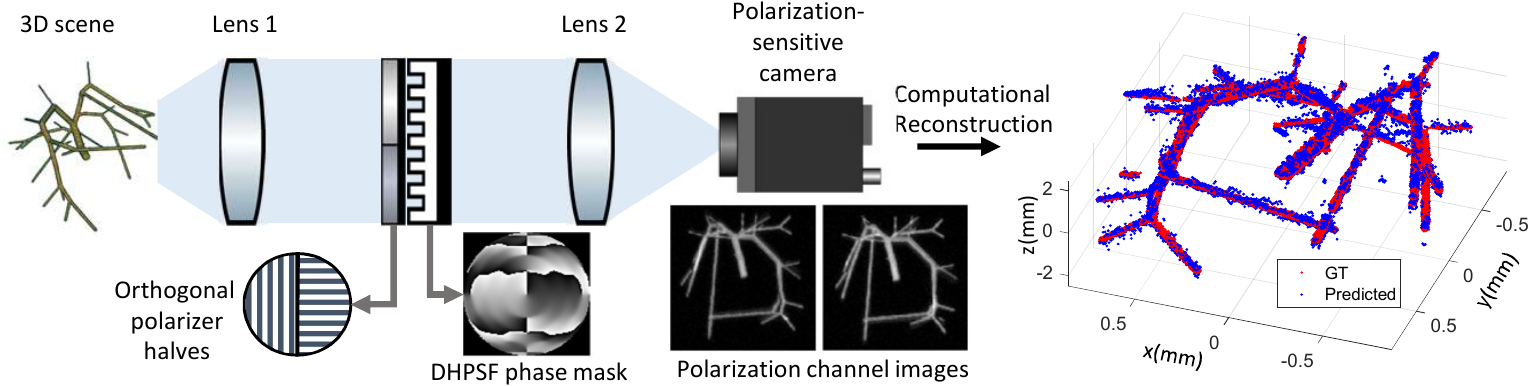}
\caption{\textbf{Polarized Spiral PSF (PS$^2$F) for single-shot 3D scene estimation.} We propose a novel monocular depth imager based on an engineered PSF. Numerous real-world scenes can be modeled as linear structures. We observe that asymmetric PSFs, such as a single lobe of the DHPSF, enable us to accurately estimate the 3D geometry of such linear structures. We leverage this observation and propose a compact polarization-based setup that produces accurate depth maps in a snapshot manner.}
\label{fig:fig1}
\end{figure*}

We propose and evaluate polarized spiral PSF (PS$^2$F), a novel engineered PSF that is well-suited for linear structures. Our key enabling observation is that asymmetric  PSFs achieve higher depth accuracy for scenes comprising  of linear structures. We achieve this asymmetry by observing that the DHPSF is comprised of two lobes that rotate about a common center; by capturing images with the individual lobes, we can reliably estimate depth. We propose a compact realization of our approach with a novel polarization-based imaging system. PS$^2$F is generated 
by combining a DHPSF mask with orthogonal linear polarizers on the two halves of the mask. We then use a polarization sensor, capable of measuring images along four polarization angles in a snapshot by using a Bayer-like tiling.
Finally, the 3D scene is estimated by solving a depth-dependent deconvolution problem. An overview of depth estimation with PS$^2$F is illustrated in Fig.~\ref{fig:fig1}. 

{\flushleft \textbf{Contributions.}} We propose a new engineered PSF for imaging linear structures and make the following contributions.
\begin{itemize}
    \item We theoretically demonstrate that the DHPSF is ill-suited for 3D imaging of linear structures by analysing its Cram\'er-Rao lower bound (CRLB).
    \item We propose a compact polarization-based setup, using a novel polarizer-phase mask encoding that is capable of better 3D estimation in a snapshot manner. 
    \item We demonstrate the advantages of PS$^2$F through several experiments with our lab prototype over a wide range of scene geometries. 
\end{itemize}

{\flushleft \textbf{Limitations.}} At its core, PS$^2$F assumes that the scene intensity is unpolarized; 
if the scene is polarized, the resultant depth estimate may not be accurate.
Our technique, like most techniques that rely on defocus blur, performs poorly for regions with little or no textures or structures.
%


\section{Related Work}\label{sec:related-work}
%
PS$^2$F draws motivation from the prior work on engineered PSFs. We discuss the relevant ones in this section.
\subsection{Mask-based PSF encoding}
Encoding PSFs by designing masks has been used in several works, including extended depth-of-field imaging~\cite{castro2004asymmetric, yang2007optimized,zammit2014extended}, 3D sensing in the macroscopic~\cite{zhou2009coded} and microscopic~\cite{pavani2009three,shechtman2014optimal} regimes, and in a lensless imaging setting~\cite{boominathan2020phlatcam}. The designed masks can be placed in front of the imaging lens or in the pupil plane of the imaging system. Earlier works include designing amplitude masks or coded apertures~\cite{levin2007image,zhou2009coded} -- such amplitude masks however lead to poor light efficiency and very coarse depth information. 
%
%
Several follow up works utilized phase-only modulation masks which include the rotating PSFs~\cite{pavani2009three,lew2011corkscrew,prasad2013rotating}, and PSFs obtained by optimizing the Fisher information~\cite{shechtman2014optimal}.
Phase masks have also been designed in an end-to-end optimization setting, both for macroscopic~\cite{wu2019phasecam3d,chang2019deep} and microscopy~\cite{nehme2020deepstorm3d,nehme2020learning} tasks.

\subsection{3D super-resolution microscopy using phase masks}
Coupled with STORM~\cite{huang2008three} and PALM~\cite{betzig2006imaging, hess2006ultra}, masks that result in rotating PSFs with depth 
have been able to perform super-resolution microscopy~\cite{pavani2009three,lew2011corkscrew} of individual fluorophore emitters to nanometer-scale accuracy. The most successful of such designs is the DHPSF~\cite{pavani2009three}, a popular choice for 3D localization of point sources~\cite{huang2008three,juette2008three,badieirostami2010three}. 3D localization performance have been further improved by designing Tetrapod PSFs~\cite{shechtman2014optimal}, which were generated by directly optimizing the Fisher information of the phase mask for $(x,y,z)$ localization of point sources. Recently, deep learning techniques have enabled an end-to-end design framework that simultaneously learn phase masks as well as neural networks  
for obtaining 3D localizations from raw captures~\cite{nehme2020deepstorm3d, nehme2020learning}.  

Of particular interest is the work by Nehme \textit{et al.}~\cite{nehme2020learning}, where a pair of masks were designed in an end-to-end learning process. The two masks are placed in two parallel $4f$ imaging channel. The mask pair, coupled with a learnt deep network, allows for high accuracy 3D localization of single fluorophores over a large depth range and for a high density of point-like emitters. Their work focuses on super-resolution microscopy and on 3D localization of \textit{point-like} emitters, 
while our result tackles extended linear structures. 

\subsection{3D sensing of linear and extended structures}
Human/animal bodies have a rich, dense 3D network of blood vessels running through them. Capturing the 3D structure of vasculature  is usually done using variants of light-sheet microscopy (LSM)~\cite{girkin2018light,di2018whole,lugo20173d}, confocal microscope (CM)~\cite{st2005confocal,kelch2015organ}, optical projection tomography~\cite{bassi2011vivo}, and optical coherence tomographic angiography~\cite{Makita:06}.
%
However, confocal and LSM methods involves scanning individual points, lines, or planes which makes the systems bulky and decreases the achievable 
imaging rate, making them non-ideal for fast dynamic 3D imaging.
%

\subsection{Depth sensing using rotating PSFs}
There is a rich body of work devoted to PSFs with lobes that rotate with defocus. 
These can be clustered into two groups based on their use of Gauss-Laguerre (GL) modes~\cite{pavani2008high} or Fresnel rings~\cite{prasad2013rotating}.

GL-based rotating PSFs~\cite{pavani2009three,lew2011corkscrew}  have been  used for super-resolving and localizing single molecules in 3D~\cite{pavani2009three,grover2011photon,gustavsson20183d, bennett2020novel}. Both  GL-based~\cite{quirin2013depth} and Fresnel-ring-based~\cite{berlich2016single,wang2017single} rotating PSFs have also been used for single-shot depth estimation of scenes. However, the first work employs a two phase-mask solution. The latter two employ a patch-based technique involving processing in the cepstrum domain. They only recover very simple scenes at low spatial and depth resolutions. Such recovery approaches do not extend to more complex geometry such as linear structures.

\subsection{Using polarization channels for PSF encoding}
There have been a few works that leverage polarization to help in 3D super-resolution imaging~\cite{roider2014axial,nehme2020learning,ikoma2021deep}. These works typically use a beam-splitter to split the incoming light into two separate $4f$ optical system channels which enabled individual channels to be encoded with a different mask. The images were then captured over two non-overlapping regions of a single sensor, or on two separate sensors. Such solutions however call for bulkier optics, and more importantly, a need for sub-pixel alignment to achieve high spatial resolution. We instead leverage snapshot polarimetric cameras that are naturally well-aligned, and hence easier to work with, and more compact.
%


\section{Polarized Spiral Point Spread Function}\label{sec:proposed-method}
We begin with a brief background on the DHPSF, an analysis of its properties, and its performance for linear/extended structures. Subsequently, we introduce our proposed PSF and its associated reconstruction technique. 

\subsection{Background: DHPSF and mask design}
This subsection outlines the background on the generation of the DHPSF \cite{pavani2009three} and other GL-based rotating PSFs \cite{lew2011corkscrew}. 

\subsubsection{Phase mask  for GL-based rotating PSFs}
Any propagating paraxial wave/beam can be expressed in terms of the orthogonal basis of GL modes, which are characterized by two integers $(n,m)$. Different combinations of the GL modes lead to varying properties of paraxial beams. In \cite{piestun2000propagation}, it was shown that beams that constituted of GL modes that are in arithmetic progression continuously rotated and scaled upon propagation. Using a beam corresponding to GL modes of (1,1), (5,3), (9,5), (13,7), it was shown in \cite{greengard2006depth} that the amplitude and phase profile of the beam at $z=0$ can be used as a mask in front of a lens (or in the Fourier plane of a $4f$ lens system) to generate a PSF that rotates with defocus. Subsequently, this mask was implemented in a phase-only manner using a variant of Gerchberg-Saxton optimization procedure with constraints included for the GL modal domain~\cite{pavani2008high}. The resulting phase function creates a PSF that rotates with changing point source depth, and is popularly known as the 
DHPSF due to its double helical structure. This PSF has been successfully used in 3D localization of fluorescent particles \cite{pavani2009three}\cite{grover2011photon}. 

\subsubsection{Properties of GL-based rotating PSFs}
Piestun \textit{et al.}~\cite{piestun2000propagation} present a theory on paraxial rotating beams using GL modes, and characterizing the resulting rotation range, rotation rates, and beam scaling rates.
Specifically, given a $4f$ system with lenses of focal length $f$,  a GL-based mask designed with beam width $w_0$ for wavelength $\lambda$, and having GL modes lying on a single line with slope $V_1$, the following can be established:
\begin{itemize}
    \item The total rotation in one direction of defocus is $(V_1\pi/2)$
    \item The angle of the rotating PSF $$\phi(z) = \phi_0 +  V_1\ \text{tan}^{-1}\left(\frac{z}{\lambda f^2/\pi w_0^2}\right)$$
    \item The PSF will rotate by $(V_1\pi/4)$ over a depth of $\frac{\lambda f^2}{\pi w_0^2}$ . 
\end{itemize}
Even though phase-only masks for rotating PSFs are generated after an optimization procedure~\cite{pavani2008high}, the above rates and ranges are still a good approximation to understand the properties of any GL-based rotating PSF.


\subsection{Challenges in imaging linear/extended structures}
\begin{figure}[!t]
\centering
\includegraphics[width=\columnwidth]{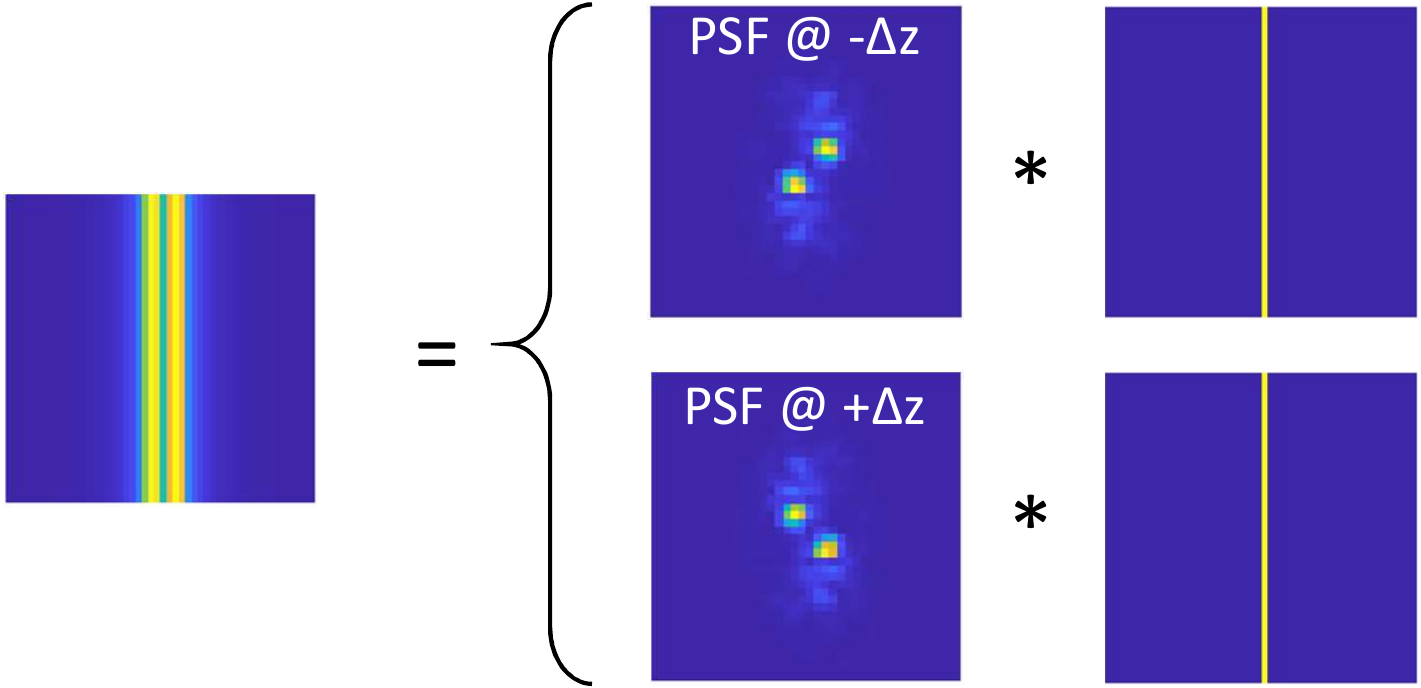}
\caption{\textbf{Ambiguities caused by DHPSF for lines}. Estimation of depth and orientation of a line  using the DHPSF is an ill-conditioned problem. There is an inherent global ambiguity in depth estimation, since for every line image, two possible candidate depths are possible.}
\label{fig:depth-ambiguity-illustration}
\end{figure}

Phase masks such as the DHPSF~\cite{pavani2009three}, or the Tetrapod PSF~\cite{shechtman2014optimal}, were designed for 3D estimation of \textit{point sources}. To image scenes with extreme fluorophore density, the concept of a scene made up of point-like emitters is not accurate. With increasing fluorophore/light source density, scenes consist of edge-like and linear structures. Using a DHPSF for recovering depth of a line then leads to an ambiguity between depth and orientation of the line. This is illustrated in Fig. \ref{fig:depth-ambiguity-illustration}. Given a line image formed using the DHPSF, there will always be a global depth ambiguity---two different depths can give rise to the same line image. This makes the DHPSF ill-suited for imaging edge/linear structures. 

\subsection{Designing a PSF for 3D lines}
%
%
A line with constant intensity blurred with an arbitrary PSF produces another line -- thereby leading to loss of information along the line direction.
In case of the DHPSF, the loss of information implies that two lines at two different depths ($-\Delta z$ and $+\Delta z$) produce the same measurement.
This ambiguity can however be resolved if we measure two separate images with each lobe of the PSF. Imaging with the individual DHPSF lobes separately will remove the global degeneracy issues (up to a $2\pi$ lobe rotation range). 
%
%
Next, we provide a theoretical analysis of the advantages to be obtained by such a separation.
For estimation from data (image(s) in our case), we can use theoretical tools to understand what ultimate precision is possible. The Fisher information and Cram\'er-Rao Lower bound (CRLB) give the best possible precision that can exist by any estimator. In the case of an image of an isolated point source, the Fisher information matrix for estimating parameters $\theta = (x,y,z)$ is given by \cite{shechtman2014optimal}:
\begin{equation}
    \label{eq:FI-point-source}
    [FI(\theta)]_{ij} = \sum_{k=1}^{N_p}\frac{1}{\mu_\theta(k) + \beta}\left(\frac{\partial \mu_\theta}{\partial \theta}\right)_i\left(\frac{\partial \mu_\theta}{\partial \theta}\right)^T_j,
\end{equation}
where $N_p$ is the number of pixels, $k$ is a variable for summing over all pixels, $T$ corresponds to the transpose operator, $\beta$ is the background Poisson noise level, and $\mu_\theta$ is the scaled PSF image for a point source at $\theta=(x,y,z)$ (scaled by the number of photons $N$). 
The diagonal elements of the inverse of the Fisher information matrix provides the CRLB for estimation of each of the parameters $\theta = (x,y,z)$. $$[\textrm{CRLB}(\theta)]_i \equiv [FI(\theta)]^{-1}_{ii} $$

We next extend the Cram\'er-Rao analysis to imaging of lines. Given a line image (or a one-sided edge image), the parameters to be estimated are the associated depth $z$ of the line/edge (at center of the patch), the associated orientation $\phi$, i.e. $\theta=(z,\phi)$. Thus, given a line image $\psi$, the CRLB estimates for depth $z$ is given by 
\begin{align}
    [FI(\theta)]_{ij} =  \sum_{k=1}^{N_p}\frac{1}{\psi_\theta(k) + \beta}\left(\frac{\partial \psi_\theta}{\partial \theta}\right)_i\left(\frac{\partial \psi_\theta}{\partial \theta}\right)^T_j \label{eq:fi}\\
    [\textrm{CRLB}(\theta)]_z \equiv [FI(\theta)]^{-1}_{zz} \label{eq:fiz}
\end{align}
The estimation of line orientation from a line path is easier than depth estimation, as orientation can be computed from the global structure of the line image. Hence, for simplicity of exposition, we focus on the analysis of the CRLB$_z$ values here. Analysis regarding the CRLB$_\phi$ values is shown in the Supplementary. Using Eqns~\ref{eq:fi}, \ref{eq:fiz} we can calculate the $\sqrt{\textrm{CRLB}_z}$ values for a given line image for all $(z,\phi)$. This calculation can be readily extended to line images captured using two separate PSFs. In such a case, the final Fisher information matrix is the sum of the individual Fisher matrices. Assuming a $4f$ system with 50~mm focal length lenses, and $N=100,000$ photons, $\beta=5$ photons/pixel, we compute the $\sqrt{\textrm{CRLB}_z}$ plots (as a function of $z$, $\phi$) for the DHPSF~\cite{pavani2008high}, the Tetrapod PSF~\cite{shechtman2014optimal}, and the individual DHPSF lobes with $25\%$ photons in each lobe. Note that $\sqrt{\textrm{CRLB}_z}$ obtained with a mask pair should be compared with a single mask having 2x SNR (2x total photons), as shown in \cite{nehme2020learning}. However, our polarizer-phase mask setup causes a $50\%$ light loss (see further in Section 3). Hence, for an appropriate comparison, we compare the individual DHPSF lobe pair, with other PSFs having $4\times$ the SNR. There are possible ways to remove this loss, with a setup demonstrated in \cite{nehme2020learning}, which will further lower the estimated $\sqrt{\textrm{CRLB}_z}$ values. 
%

\begin{figure}[!t]
\centering
\includegraphics[width=\columnwidth]{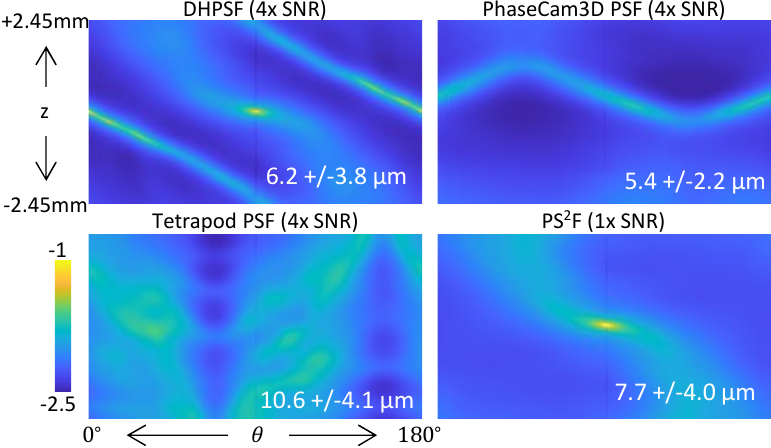}
\caption{\textbf{CRLB comparison for various phase masks.} The figure shows $\sqrt{CRLB_z}$ log-intensity (log-mm) plots for line images, as a function of line depth $(z)$ and line orientation $(\theta)$. PS$^2$F achieves a comparable CRLB over the whole range of depth and orientations with fewer peaks, even with 4x lower SNR. In contrast, competing approaches have curves with large CRLB values, which introduce ambiguities in the depth estimate.}
\label{fig:crlb-plots}
\end{figure}

Fig.\ \ref{fig:crlb-plots} illustrates the $\sqrt{\textrm{CRLB}_z}$ values (log-scale) for a given line at depth $z$ and having orientation $\phi$. The mean and standard deviation of the $\sqrt{\textrm{CRLB}_z}$ values are also shown in the individual insets. The DHPSF plot in Fig. \ref{fig:crlb-plots} shows  several peaks with large magnitude. This occurs at specific $(z,\phi)$ where the line connecting the two lobes of the PSF and the scene line are either perpendicular or parallel to each other. 
The ability to discern lines at slightly differing depths is greatly reduced at these points due to symmetry. For the Tetrapod PSF~\cite{shechtman2014optimal} and the PhaseCam3D PSF~\cite{wu2019phasecam3d}, the plots in Fig. \ref{fig:crlb-plots} also show more peaks and ridges, but with lower magnitude compared to the DHPSF. For the PSF pair of the individual DHPSF lobes, we see that most of the peaks are removed, except for a central peak in the plot. The peak corresponds to the case when the PSF is in focus and the the line orientation is perpendicular to the PSF lobes. This central peak does not affect the effectiveness of separating lobes, because it is a narrow peak over a only small area in the parameter space of $(z,\phi)$. The average $\sqrt{\textrm{CRLB}_z}$ values for the individual DHPSF lobe pair is similar to the DHPSF with 4x SNR. Thus, with 4x lower SNR, the PS$^2$F is able to achieve a comparable $\sqrt{\textrm{CRLB}_z}$ value with other PSFs, and does not show consistent peaks and ridges as seen in the plots. Note that CRLB calculations only provide insight about local ambiguities or precision levels. However, we also observe that a PSF pair created out of separating the DHPSF lobes removes global ambiguities as well. Thus overall, a PSF pair created from the individual DHPSF lobes is better for estimating depths of line/edge patches.  

\subsection{Using polarizers to separate out the DHPSF lobes in a single $4f$ optical system}
\begin{figure}[!t]
\centering
\includegraphics[width=\columnwidth]{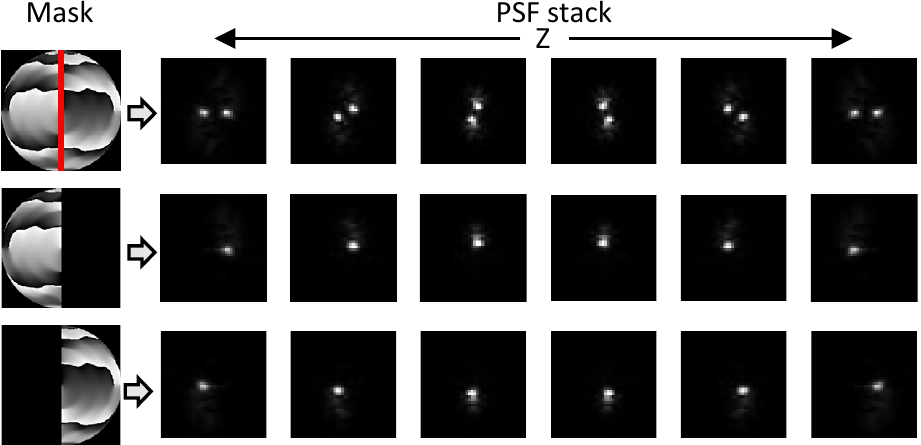}
\caption{\textbf{Effect of separating the phase mask.} The phase mask for generating the DHPSF has a special property that separating the phase profile along the highlighted red line yields a separation of the two lobes. We leverage this to implement a compact realization of Polarized Spiral PSF (PS$^2$F) using a polarization camera sensor and two polarizers in the pupil plane.}
\label{fig:psf-partition-illustration}
\end{figure}

Imaging with a PSF pair is possible by having two separate parallel $4f$ imaging systems with separate sensors, as demonstrated in \cite{nehme2020learning}.
However, the DHPSF mask (and the Fresnel ring-based two-lobe masks) enjoy the property of PSF lobe separability. Partitioning such a mask into two halves along a particular axis, and allowing light through only one half leads to the creation of only a single rotating PSF lobe, as depicted in Fig.~\ref{fig:psf-partition-illustration}. The primary reason is that the light falling on one half is modulated to form one lobe, while the light falling on the other half is modulated to form the other lobe. Such a partitioning leads to partitioning of the DHPSF into two separate, distinct lobes.  

The separability of DHPSF lobes can be leveraged to build a compact imaging system. In PS$^2$F, we add two linear polarizers on each mask half, one oriented along $s$-polarization and the other oriented along $p$-polarization.
This ensures that the $s$-polarization component is modulated using one mask half, and consequently the $p$-polarized component is modulated using the other mask half. The PSF created out of such a polarizer-phase mask encoding contains two lobes, but with each lobe being in a different polarization (either $s$ or $p$). This addition of polarizers causes a $50\%$ light loss, which was taken into consideration while estimating the $\sqrt{\textrm{CRLB}_z}$ values. Such a \textit{polarized spiral PSF} can be captured efficiently using a polarized camera sensor. A polarized camera sensor contains a 2D array of pixels with a Bayer pattern that has 90, 45, 135, 0 degree polarizers. After addition of the polarizers in the mask-plane, the orthogonally polarized DHPSF lobes can be imaged in their respective channels of a polarization sensor. This allows for a compact $4f$ system, allowing the imaging of a PSF pair using just a single $4f$ system channel, making the proposed method a single-shot, single-sensor method. 
\subsection{Imaging model and reconstruction procedure}
We assume that the light from the scene is unpolarized in our imaging model. The imaging of a 3D scene with a depth-dependent PSF can be approximated as a sum of 2D convolutions between the depth-dependent PSF and the per-plane scene intensity: 
\begin{equation}
    \label{eq:imaging-model}
    I_c(x,y) = \sum_{z=z_{start}}^{z_{end}} h_c(x,y;z) * s(x,y,z)
\end{equation}
where $I_c(x,y)$ is the image intensity at $(x,y)$ in polarization channel $c$, $h_c(x,y;z)$ is the 2D PSF corresponding to depth $z$ and polarization channel $c$, and $s(x,y,z)$ is the scene intensity at point $(x,y,z)$.

The goal of 3D reconstruction is to estimate a 3D matrix $s(x,y,z)$ from the noisy measurements of $I(x,y)$. 
There are numerous ways to solve this problem including regularized least-squares~\cite{boominathan2020phlatcam,yanny2020miniscope3d, xue2020single}, data learning-based~\cite{wu2019phasecam3d,chang2019deep}, and more recently, ones based on deep network-based regularizer~\cite{zhang20203d}.
To keep the exposition simple, we demonstrate recovery by modelling the 3D reconstruction problem as a regularized least squares optimization problem.
%

Specifically, we formulate the 3D scene estimation problem as:
\begin{equation}
    \label{eq:optim_func}
    \underset{\textbf{x}}{\text{argmin}} \left\lVert \begin{bmatrix}I_0\\I_{90}\end{bmatrix} - S\begin{bmatrix}H_0\\H_{90}\end{bmatrix}\textbf{x}\right\rVert^2_2 + \lambda_{TV}\lVert\Psi(\textbf{x})\rVert_1 + \lambda_{L1}\lVert \textbf{x} \rVert_1
\end{equation}
where $\textbf{x}$ is a 3D matrix of scene intensities, $H_0, H_{90}$ are the depth-specific PSF operators corresponding to the individual two lobes that are polarized to $0^\circ$ and $90^\circ$ states respectively, and $I_0, I_{90}$ are captured images in the $0^\circ$ and $90^\circ$  polarization channels respectively. $S$ is the summing operator that sums across the depth channels of $H_0\textbf{x}$ and $H_{90}\textbf{x}$ separately. We employ TV and L1 regularizers as scene priors, with $\lambda_{TV}, \lambda_{L1}$ as hyperparameters to control their regularization effects respectively. 
We solve the optimization problem using autograd functionality in PyTorch and the Adam optimizer~\cite{kingma2014adam} to leverage the speed of graphical processing units (GPUs). 

A key requirement for our technique is that the scene emits unpolarized light. This is required so as to ensure that both the lobes have produce PSFs with similar intensity levels that we can calibrate a priori. When the incident light is polarized, the blur kernels in Eq.~\ref{eq:imaging-model} will have an unknown scaling, which leads to a model mismatch;  in an extreme scenario, we lose all the information in one of the lobes if the polarization angle of the incident light is  orthogonal to the corresponding polarizer in the pupil plane. While this is a limitation of our technique, we largely encountered unpolarized light in our intended application of fluorescence microscopy.


\section{Simulations}\label{sec:simulations}
%

\begin{figure*}[!t]
\centering
\includegraphics[width=\textwidth]{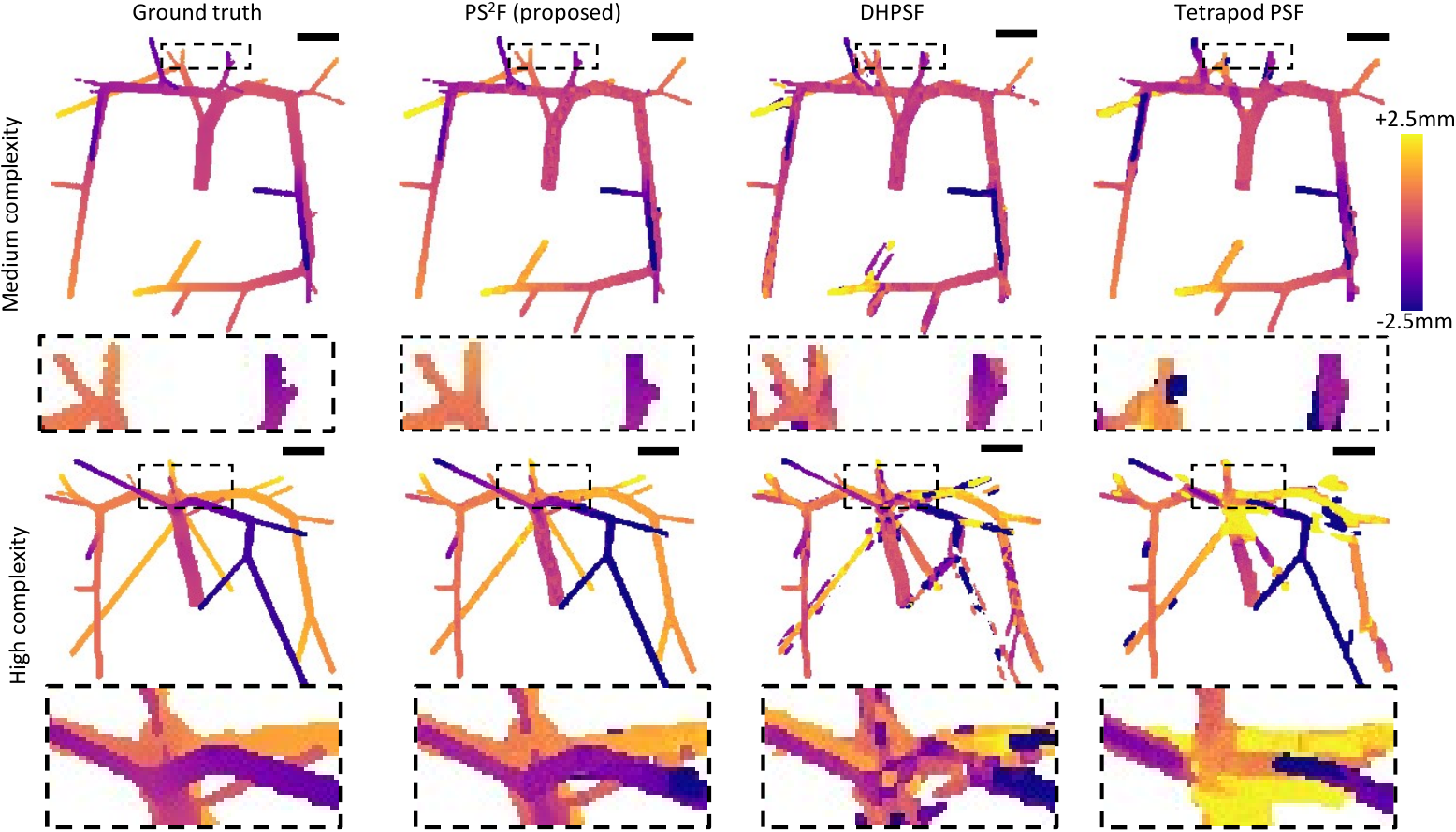}
\caption{\textbf{Simulations on vascusynth dataset with varying complexity.} The figure shows reconstructions with various PSF designs on scenes with medium complexity (top row) and high complexity (bottom row). PS$^2$F is superior in depth estimation, especially in regions where there is higher vasculature (scene) complexity (see insets). Scale bars indicate $0.25$~mm.}
\label{fig:vasc-scenes-dmaps}
\end{figure*}

To evaluate the performance of our proposed PS$^2$F, we perform extensive simulations on the \href{http://vascusynth.cs.sfu.ca/Welcome.html}{VascuSynth 2013} dataset~\cite{hamarneh2010vascusynth, jassi2011vascusynth}, which provides 10 simulated 3D volumes of vascular trees for each of 12 levels of complexity. Scene complexity is set by the number of bifurcations in the vascular tree (ranging from 1 to 56 in steps of 5). 
\begin{figure}[!t]
\centering
\includegraphics[width=\columnwidth]{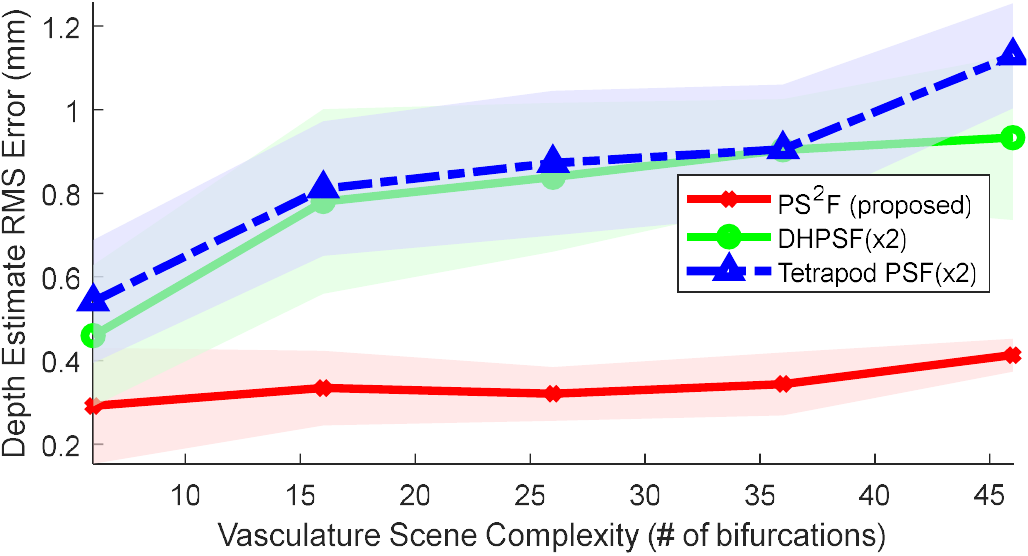}
\caption{\textbf{Accuracy vs.\ scene complexity}. Depth estimation performance, for the proposed PS$^2$F and other comparing PSFs (averaged over 10 scenes for each complexity level, shaded regions show variance across the 10 scenes). This demonstrates the suitability of PS$^2$F, and that the separation of the DHPSF lobes helps in 3D reconstruction.}
\label{fig:sim-compare}
\end{figure}
Each 3D vasculature volume in the dataset consists of $101\times101\times101$ voxels. These volumes were converted into 3D scenes of size $256\times256\times256$ through interpolation methods. This new volume was assumed to cover a volume of $1.76\times1.76\times5.00$ mm$^3$ with each voxel occupying a  $6.9\times6.9\times19.5~\mu$m$^3$.

We further assume a $4f$ system with both lenses having focal lengths $f=50$~mm. We render images of the vascular structures using several different masks of diameter $3$~mm in the Fourier plane. We assume that the light from the vasculature is monochromatic of wavelength $\lambda=532$~nm. We render images assuming occlusion of vasculature and that light is received only from the surface of the vasculature. We assume that the center of the 3D scene is $f=50$~mm away from the first lens, thus, the scene spans a defocus range of [$-2.5$~mm, $+2.5$~mm].  The masks (and their corresponding PSFs) used in rendering the simulated images were:
\begin{itemize}
    \item DHPSF: We optimize for a GL-based DHPSF mask (following procedure in~\cite{pavani2008high}), based on the rotating paraxial beam with $w_0=0.4$~mm and having GL modes corresponding to $(1,1), (3,5), (5,9), (7,13)$. Full 180 degree lobe rotation was achievable over $2\frac{\lambda f^2}{\pi w_0^2}=5.3$~mm. 
    \item PS$^2$F: Using the above mask, we construct an equivalent mask pair corresponding to PS$^2$F by partitioning the mask into two halves appropriately. 
    \item Tetrapod PSF: We obtain a Tetrapod PSF mask designed for $550$~nm wavelength from~\cite{gustavsson20183d} and repurpose it for imaging over our specified $5$~mm depth range. 
\end{itemize}
We then added Poisson and Gaussian read-out noise to the final measurements. For an appropriate comparison, we perform 3D reconstructions with individual single lobes (PS$^2$F), two-shot capture with DHPSF, and a two-shot capture using the Tetrapod PSF. To accurately account for the $50\%$ light loss in the PS$^2$F case, we also render out images with half the signal level as compared to the signal levels in DHPSF and Tetrapod PSF case. 
After obtaining the 3D volumetric estimate, we filter out points whose sum across the z-stack is lower than a fixed threshold. We  then estimate depth using the index corresponding to a maximum-intensity-projection (MIP) of the 3D estimate. 

\subsection{Comparison with DHPSF and Tetrapod PSF}
Fig. \ref{fig:sim-compare} shows the depth estimation performance for the proposed PS$^2$F, DHPSF, and the Tetrapod PSF. We compare the PSFs depth estimation using the Mean Absolute Error (MAE), Root Mean Square Error (RMSE), and the multi-scale Structural Similarity Index Measure~\cite{wang2003multiscale} (MS-SSIM) metrics. The entire set of results can be seen in the Supplementary. The proposed PS$^2$F method performs $\sim2\times$ better in terms of the RMS error for depth estimation. The depth estimation performance improvement is even higher ($\sim3\times$) for vasculature with more bifurcations, i.e. greater complexity. Depth estimation results for two example vasculature scenes can be seen in Fig. \ref{fig:vasc-scenes-dmaps}. In Fig. \ref{fig:vasc-scenes-dmaps} (top row), the vasculature scene is not very complex. The depth map estimation results are fairly similar for all the three PSFs (PS$^2$F, DHPSF, Tetrapod PSF), but the zoom-in insets highlight the errors seen in the DHPSF and Tetrapod PSF depth estimation. In Fig. \ref{fig:vasc-scenes-dmaps} (bottom row), the scene has greater complexity, and the PS$^2$F produces much better depth estimation than the other two PSFs. The zoom-in insets especially highlight the same. 

\begin{figure}[!t]
\centering
 \centering
 \includegraphics[width=\columnwidth]{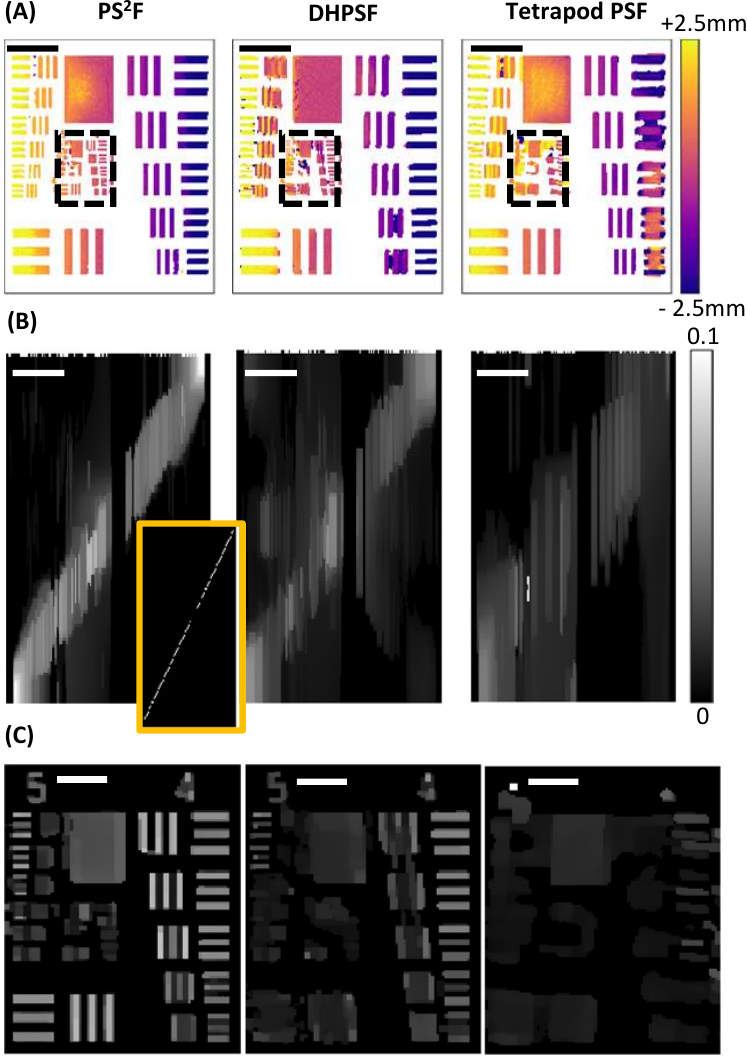}
 \caption{\textbf{Simulated reconstruction results of a USAF target at a skew angle.} (A) Shows the depth map estimates for PS$^2$F, DHPSF, and Tetrapod PSF respectively. The XZ maximum intensity projections are shown in (B) in their respective columns, with ground truth shown in the inset. (C) shows zoom-in insets of depth maps (after thresholding to remove small artifacts) of highlighted regions in (A). PS$^2$F has a better reconstruction in lateral and axial dimensions as compared to the DHPSF and the Tetrapod PSF. Scale bars for (A), (B), (C) indicate $0.5$~mm, $0.5$~mm, and $0.1$~mm respectively.}
\label{fig:sims-usaf}
\end{figure}
\subsection{Resolution performance}
Furthermore, we analyse the reconstruction resolution performance by reconstructing a simulated skew USAF target. We assume a USAF target places at the focus plane, but tilted with respect to the optical axis so that its  depth changing horizontally across the target; we render using the same three masks in the same $4f$ optical system as above. The results are shown in Fig.\ \ref{fig:sims-usaf}. The proposed PS$^2$F generates a better reconstruction than the DHPSF and Tetrapod PSF reconstructions, both in lateral and axial dimensions. As seen from the XZ MIP plots in Fig.\ \ref{fig:sims-usaf}, the PS$^2$F and Tetrapod PSF reconstructions are able to obtain the skew angle of the target, with the former being much better. The DHPSF fails to get the skew angle, due to the global depth ambiguity issues explained in Section \ref{sec:proposed-method}. In the lateral dimensions, the Tetrapod PSF is unable to reconstruct Group 4 and 5. This could be attributed to the fact that the PSF has a larger support. On the other hand, the PS$^2$F and DHPSF have concentrated lobes, and thus are able to reconstruct elements of Group (5,3) (12.40~$\mu$m linewidth) and (5,2) (13.92~$\mu$m linewidth) respectively.  

The obtained $xy$-resolution can be theoretically justified. The separation of lobes in the PS$^2$F method allows for resolution of finer scene features due to the compactness of the PSF. The resolution will be roughly determined by the width of the individual PSF lobes. Fitting a 2D Gaussian to each individual lobe, the average $2\sigma$-value for the Gaussian fit is $\sim1.74~\text{pixels} = 12~\mu$m in the object space. This readily matches with the fact that the PS$^2$F can resolve elements of Group (5,3), which has line elements that are spaced at $12.40~\mu$m distance. The diffraction-limited resolution of an Airy disk PSF for the same optical system configuration is given by $1.22\lambda f/D = 10.81~\mu$m, indicating that the $xy$-resolution obtained by the PS$^2$F is close to the diffraction-limited resolution ($1.15\times$) as well. 

For PS$^2$F, we also analysed the spread of reconstruction signal across the z-channel. For pixels with a significant signal level, we fit a 1D Gaussian curve to estimate this spread. The median $2\sigma$-value across z-channel obtained was $2\times304=608~\mu$m, which is an estimate of the axial resolution performance of the proposed PS$^2$F. Note that the diffraction-limited axial performance of an Airy disk PSF for the same optical system configuration is given by $4\lambda(f/D)^2 = 591~\mu$m. We are able to achieve $\sim1.03\times$ the diffraction-limited axial resolution with the proposed PS$^2$F.  


\section{Experimental Results}\label{sec:experiments}
\begin{figure}[!t]
\centering
\includegraphics[width=\columnwidth]{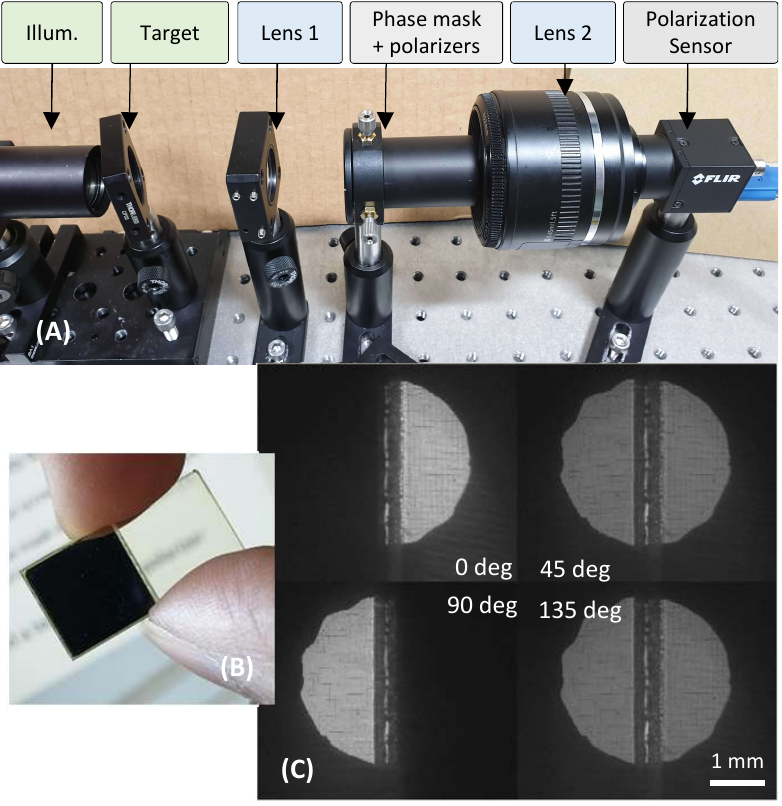}
\caption{\textbf{Lab prototype.} (A) Our lab prototype is a $4f$ system consists of a polarization sensor (FLIR BFS-U3-51S5P-C) equipped with a $50$~mm F-mount lens. The phase mask and polarizers are on an XY stage. An achromat (AC254-050-A) forms the other half of the $4f$ system. We used an additional z-axis motorized translation stage to capture the PSF stack. (B) Two orthogonal polarizer pieces. (C) Images of the polarizers and phase mask with a camera focused on the pupil-plane.}
\label{fig:4f-setup}
\end{figure}

We perform real-world imaging with PS$^2$F to verify our simulations. In this section we detail out the experiments and compare the PS$^2$F with the corresponding DHPSF.


\subsection{Imaging setup description}
We built a $4f$ imaging setup with the first lens being an achromatic lens with focal length $50$~mm, and the second lens being a Canon camera lens with effective focal length $=50$~mm. The polarizer-phase mask is added to the pupil plane of the $4f$ system. Fig. \ref{fig:4f-setup} shows the prototype. 

\subsubsection{Polarizer-Mask design}
We fabricated a $3$~mm diameter DHPSF mask using a 3-D printer employing two-photon lithography. See Supplementary for more details on fabrication. The fabricated mask was the same mask used in the Section \ref{sec:simulations} to perform simulations of the VascuSynth dataset. To create a single-mask, single-sensor design for the PS$^2$F as outlined in Section \ref{sec:proposed-method}, two rectangular pieces of thin-film polarizers were laser cut and put side-by-side in a 3D printed holder. Both the polarizer pieces were laser cut such that their polarization orientation was $0\deg$ and $90\deg$ respectively. The fabricated DHPSF mask was attached to the back side of the 3D printed holder, carefully aligning the two mask halves w.r.t. to the polarizer edges under a bright-field 4x microscope. This polarizer-mask aperture is illustrated in Fig. \ref{fig:4f-setup}(B),(C). Further details can be found in the Supplementary figures.
\subsubsection{PSF capture and calibration}
\begin{figure}[!t]
\centering
\includegraphics[width=\columnwidth]{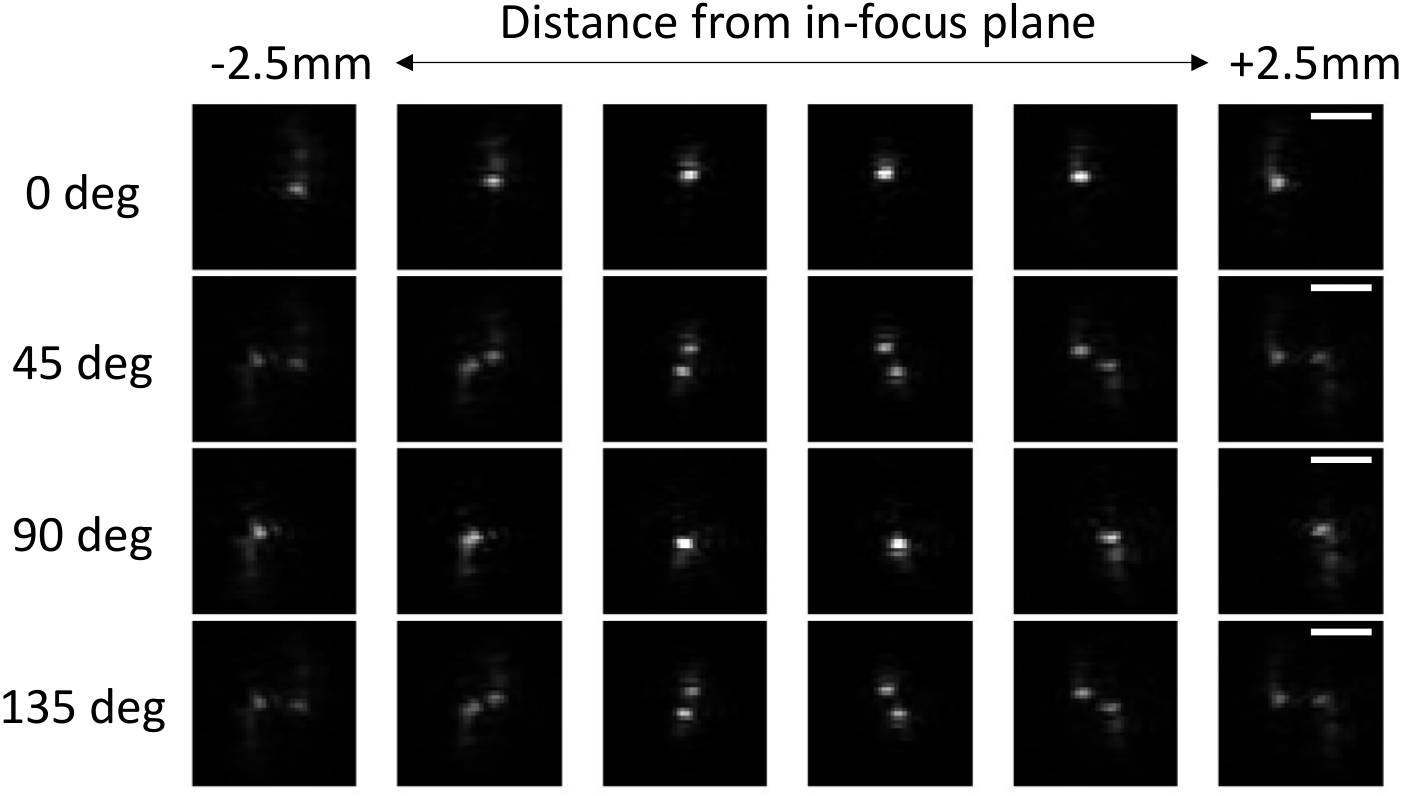}
\caption{\textbf{PSF stacks captured with a lab prototype.} We achieved high quality separation between the two lobes in 0 and 90 degree images with our setup. We noticed no significant effect of any misalignment between the phase mask and the two polarizer halves. Scale bars indicate $100~\mu$m.}
\label{fig:psfstack-vis}
\end{figure}

To reconstruct 3D scenes, we use an experimentally captured PSF stack for the PS$^2$F. Using the polarization sensor, we image a $5~\mu$m pinhole over a range of [$-2.5$~mm, $+2.5$~mm] on either side of the focus plane. The experimentally-captured PSFs can be seen in Fig.~\ref{fig:psfstack-vis}. 
\subsection{Imaging experiments}
With the fabricated mask and optical setup, we perform a series of experiments with 3D skew planar targets (using transmissive illumination) and with 3D fluorescence targets. We image using a polarization camera sensor (FLIR BFS-U3-51S5P-C) to obtain the individual PS$^2$F lobes in two polarization channels. By averaging across the Bayer pattern, we can readily obtain the image of the scene as imaged by a DHPSF, which we use for comparison purposes.

\begin{figure*}[!t]
\centering
\includegraphics[width=\textwidth]{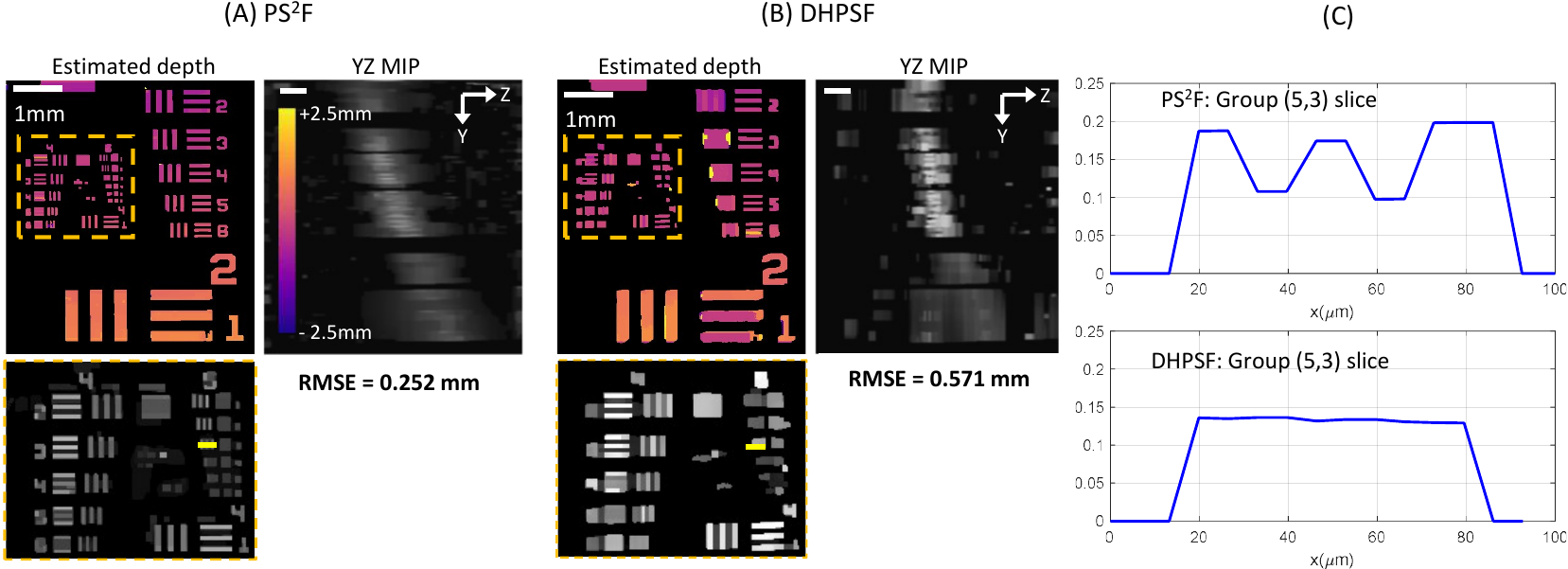}
\caption{\textbf{Skew USAF target reconstruction.} Showing reconstructed depth map, YZ maximum intensity projection, zoom-in inset of highlighted ROI for (A) PS$^2$F, and for (B) DHPSF. Column (C) shows slice plots for marked yellow lines in the zoom-in insets, corresponding to Group (5,3) elements for PS$^2$F (top plot) and DHPSF (bottom plot).  We were able to resolve up to Group (5,3) elements (12.40 $\mu$m linewidth) using the PS$^2$F, whereas only up to Group (4,5) elements (19.69 $\mu$m linewidth) using the DHPSF. Text below YZ MIP plots indicate the depth RMS error obtained for reconstruction when compared to ground truth, showing that the PS$^2$F performs $\sim2\times$ better. All scale bars indicate $1$~mm.}
\label{fig:expt-usaf}
\end{figure*}

\begin{figure*}[!t]
\centering
\includegraphics[width=\textwidth]{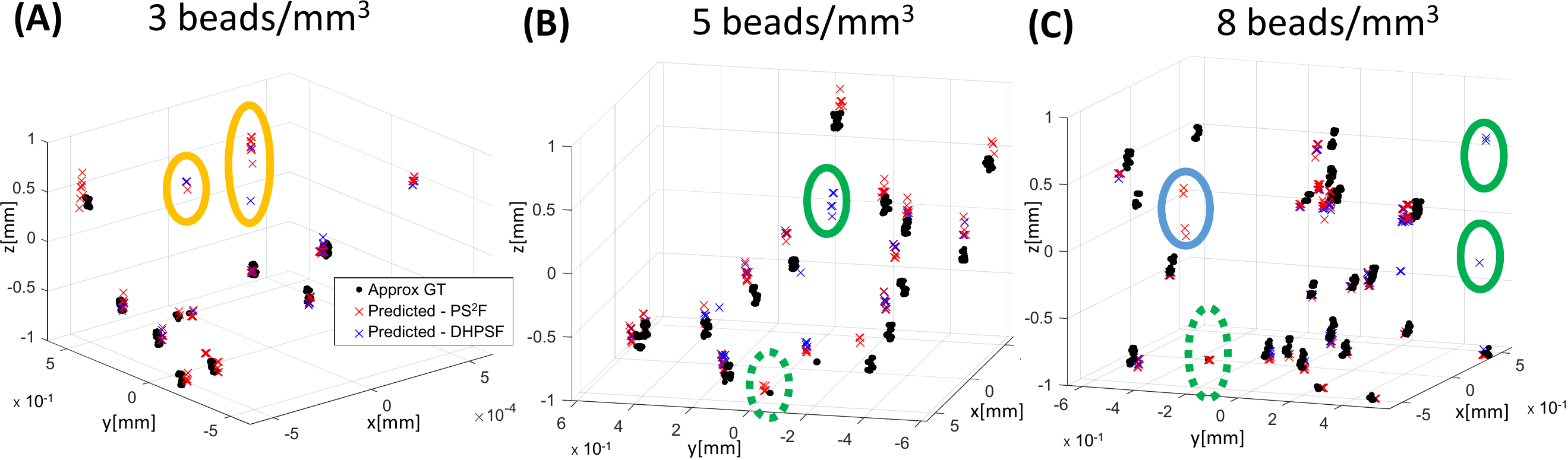}
\caption{\textbf{3D reconstructed fluorescent bead volumes.} The scatter plots show 3D reconstructed views with PS$^2$F and the DHPSF for bead densities approximately corresponding to (A) 3 beads/mm$^3$, (B) 5 beads/mm$^3$, and (C) 8 beads/mm$^3$. We captured ground truth information with a semi-aligned confocal ground truth (hence an approximate ground truth). Highlighted regions indicate various errors: incorrect estimation by DHPSF \textcolor{mygreen}{(solid green)}, missing estimation by DHPSF \textcolor{mygreen}{(dashed green)}, incorrect estimation by PS$^2$F \textcolor{myblue}{(solid blue)}, and incorrect estimation by both PSFs \textcolor{myyellow}{(solid yellow)}. Axes values are in units of $mm$.}
\label{fig:expt-fluorescence}
\end{figure*}
\begin{figure*}[!t]
    \centering
    \includegraphics[width=\textwidth]{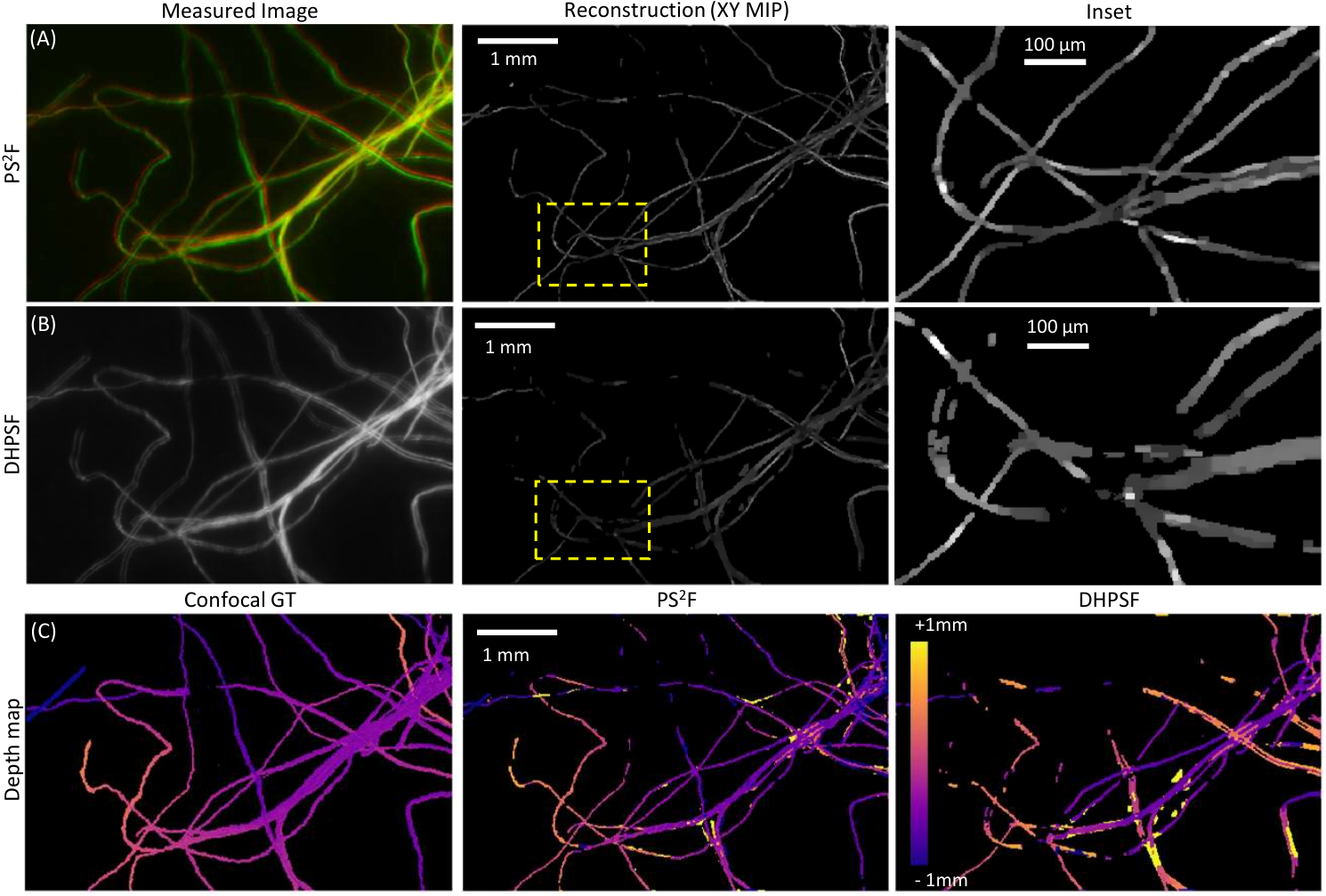}
    \caption{\textbf{Experimental results on fluorescent cotton strands}. (A) PS$^2$F and (B) DHPSF. First column in (A), (B) show the captured scene images (polarization channels in the PS$^2$F image are shown in red and green colors). Second column in (A), (B) shows the reconstructions. Highlight ROIs (in yellow) are shown in the last column of (A), (B) -- depicting the inaccuracy of DHPSF, as well as the ability of PS$^2$F to reconstruct scenes with complex linear structures. In the last row (C), columns (from left-to-right) show color-coded depth maps obtained from confocal imaging (treated as ground truth), PS$^2$F reconstructions, and DHPSF reconstructions respectively. Note the confocal GT depth map is an approximate depth map, due to imperfect registration between GT and the imaged volumes.}
    \label{fig:expt=vasc}
\end{figure*}
\subsubsection{3D skew planar targets}
We imaged a USAF planar target placed at a skew angle, and then performed single-shot 3D reconstruction of the target. Furthermore, we obtain a focal stack of the target so as to obtain ground truth depth map. The ground truth depth map was obtained from the focal stack using the Shape-from-focus method~\cite{peruz}. For the FOV being imaged, the relative intensity of each of the PS$^2$F lobes changed across the FOV, causing a spatial variance in the PSF. This was readily accounted for by adding a per-pixel, per-channel weight term to the 3D scene variable \textbf{x} in Eq. \ref{eq:optim_func}, and jointly optimizing for the 3D scene \textbf{x} and the weights. The results are depicted in Fig. \ref{fig:expt-usaf}. After modelling of the relative strengths of the PS$^2$F lobes as variables to be estimated during the optimization, we achieved an accurate reconstruction result  matching the simulation results in terms of resolving Group (5,3) elements for the PS$^2$F. The YZ maximum intensity projections demonstrate that the skew plane of the USAF has been correctly estimated using the PS$^2$F, as compared to using the DHPSF. When judged against the ground truth, we see a $\sim50\%$ lower RMS error achieved with the PS$^2$F as compared to the DHPSF. 

\subsubsection{3D fluorescence imaging}
We, next, imaged a set of prepared fluorescent samples, consisting of $10~\mu$m fluorescent beads in a PDMS substrate (for more information, see Supplementary). The bead concentration was varied from approximately $3$ to $8$ beads/mm$^3$. The resulting samples were $\sim2$~mm in thickness. We imaged these samples using the same mask as before, but in a $4f$ optical system using lenses of focal length $30$~mm and $50$~mm resulting in a $1.6\times$ magnification. 
We reconstructed using the same optimization algorithm as specified in Eq. \ref{eq:optim_func}, and compared against ground truth data obtained with a confocal microscope. 
The results are shown in Fig.\ \ref{fig:expt-fluorescence}. Bead localization performance using the DHPSF seems to get relatively worse as the bead concentration increases. In contrast, PS$^2$F enabled accurate results with increasing bead concentration. It is important to note that in this experiment, we use an optimization objective to perform bead reconstruction. More sophisticated and accurate methods (such as likelihood-based methods) exist for reconstruction of beads or point-like objects. Moreover, the ground truth (GT) data was captured using confocal imaging method. However, due to the complexity of matching/registering imaging volumes, we only obtain an approximated alignment or approximate registration between the GT and the imaging volume captured.

Using the same optical setup as above, we also image fluorescent cotton strands suspended in a PDMS sample (see Supplementary for more information about sample preparation). Fig.~\ref{fig:expt=vasc} shows the results for the same. PS$^2$F method is better able to capture the 3D structure of the fluorescent strands - especially in parts with highly complex geometry. Imaging with PS$^2$F results in sharper reconstructions, which helps in reconstructing the fine, complex structure of the strands. The estimated depth map from the PS$^2$F is also more accurate as compared to the DHPSF case.  

\section{Conclusion}\label{sec:conclusions}

We presented PS$^2$F for single shot, monocular depth estimation of extended (linear) structures where we separated the two lobes of the DHPSF into the two orthogonal states of polarization.
This separation enabled us to remove ambiguities when estimating depth of line segments by breaking the inherent symmetry of DHPSF with respect to depth.
The snapshot capabilities of PS$^2$F will enable faster microscopic imaging, including high resolution light sheet microscopy at real-time rates with fewer captures.
We designed and demonstrated a compact physical realization of PS$^2$F with a single, polarization-sensitive camera, and showed that PS$^2$F results in $2\times$ or higher accuracy compared to current state-of-the-art phase masks such as DHPSF and the Tetrapod PSF.
We believe that our approach of leveraging polarization multiplexed phase masks combined with polarization-sensitive cameras opens new avenues for high resolution snapshot depth imaging at micro and macro scales.

PS$^2$F is inherently designed for extended/linear structures, and imaging such structures will involve high photon counts than when imaging single molecules or point-like emitters. Hence, we do not specifically deal with a low-photon count scenario in this work.
Nevertheless, PS$^2$F can be adapted to low-photon count scenarios by using advanced recovery techniques including deep network-based approaches. 
Improvements such as modeling spatially varying PSF, as well as accounting for polarized input light will further improve PS$^2$F results.




\ifpeerreview \else
\section*{Acknowledgments}
The authors would like to thank Dong Yan for his help in preparing the fluorescent bead and fluorescent strands samples. This work was supported by NSF awards IIS-1730574, IIS-1730147, CCF-1652569, IIS-1652633, and EEC-1648451. A.-K.G. acknowledges partial financial support from the National Institute of General Medical Sciences of the National Institutes of Health (Grant No. R00GM134187), the Welch Foundation (Grant No. C-2064-20210327), and startup funds from the Cancer Prevention and Research Institute of Texas (Grant No. RR200025).
\fi

\bibliographystyle{IEEEtranN}
\bibliography{references}

\begin{thebibliography}{54}
\providecommand{\natexlab}[1]{#1}
\providecommand{\url}[1]{#1}
\csname url@samestyle\endcsname
\providecommand{\newblock}{\relax}
\providecommand{\bibinfo}[2]{#2}
\providecommand{\BIBentrySTDinterwordspacing}{\spaceskip=0pt\relax}
\providecommand{\BIBentryALTinterwordstretchfactor}{4}
\providecommand{\BIBentryALTinterwordspacing}{\spaceskip=\fontdimen2\font plus
\BIBentryALTinterwordstretchfactor\fontdimen3\font minus
  \fontdimen4\font\relax}
\providecommand{\BIBforeignlanguage}[2]{{%
\expandafter\ifx\csname l@#1\endcsname\relax
\typeout{** WARNING: IEEEtranN.bst: No hyphenation pattern has been}%
\typeout{** loaded for the language `#1'. Using the pattern for}%
\typeout{** the default language instead.}%
\else
\language=\csname l@#1\endcsname
\fi
#2}}
\providecommand{\BIBdecl}{\relax}
\BIBdecl

\bibitem[Fischer et~al.(2011)Fischer, Wu, Kanchanawong, Shroff, and
  Waterman]{fischer2011microscopy}
R.~S. Fischer, Y.~Wu, P.~Kanchanawong, H.~Shroff, and C.~M. Waterman,
  ``Microscopy in 3d: a biologist's toolbox,'' \emph{Trends in Cell Biology},
  vol.~21, no.~12, pp. 682--691, 2011.

\bibitem[Arnold et~al.(2019)Arnold, Al-Jarrah, Dianati, Fallah, Oxtoby, and
  Mouzakitis]{arnold2019survey}
E.~Arnold, O.~Y. Al-Jarrah, M.~Dianati, S.~Fallah, D.~Oxtoby, and
  A.~Mouzakitis, ``A survey on 3d object detection methods for autonomous
  driving applications,'' \emph{IEEE Trans. Intelligent Transportation
  Systems}, vol.~20, no.~10, pp. 3782--3795, 2019.

\bibitem[Reiter et~al.(2014)Reiter, Sigaras, Fowler, and
  Allen]{reiter2014surgical}
A.~Reiter, A.~Sigaras, D.~Fowler, and P.~K. Allen, ``Surgical structured light
  for 3d minimally invasive surgical imaging,'' in \emph{IEEE Intl. Conf.
  Intelligent Robots and Systems}, 2014.

\bibitem[Ping-Sing and Shah(1994)]{ping1994shape}
T.~Ping-Sing and M.~Shah, ``Shape from shading using linear approximation,''
  \emph{Image and Vision Computing}, vol.~12, no.~8, pp. 487--498, 1994.

\bibitem[Levin et~al.(2007)Levin, Fergus, Durand, and Freeman]{levin2007image}
A.~Levin, R.~Fergus, F.~Durand, and W.~T. Freeman, ``Image and depth from a
  conventional camera with a coded aperture,'' \emph{ACM Trans. Graphics},
  vol.~26, no.~3, pp. 70--es, 2007.

\bibitem[Zhou et~al.(2009)Zhou, Lin, and Nayar]{zhou2009coded}
C.~Zhou, S.~Lin, and S.~Nayar, ``Coded aperture pairs for depth from defocus,''
  in \emph{IEEE Intl. Conf. Computer Vision (ICCV)}, 2009.

\bibitem[Pavani et~al.(2009)Pavani, Greengard, and Piestun]{pavani2009three}
S.~R.~P. Pavani, A.~Greengard, and R.~Piestun, ``Three-dimensional localization
  with nanometer accuracy using a detector-limited double-helix point spread
  function system,'' \emph{Appl. Physics Letters}, vol.~95, no.~2, p. 021103,
  2009.

\bibitem[Greengard et~al.(2006)Greengard, Schechner, and
  Piestun]{greengard2006depth}
A.~Greengard, Y.~Y. Schechner, and R.~Piestun, ``Depth from diffracted
  rotation,'' \emph{Optics Letters}, vol.~31, no.~2, pp. 181--183, 2006.

\bibitem[Pavani and Piestun(2008)]{pavani2008high}
S.~R.~P. Pavani and R.~Piestun, ``High-efficiency rotating point spread
  functions,'' \emph{Optics Express}, vol.~16, no.~5, pp. 3484--3489, 2008.

\bibitem[Shechtman et~al.(2014)Shechtman, Sahl, Backer, and
  Moerner]{shechtman2014optimal}
Y.~Shechtman, S.~J. Sahl, A.~S. Backer, and W.~E. Moerner, ``Optimal point
  spread function design for 3d imaging,'' \emph{Physical Review Letters}, vol.
  113, no.~13, p. 133902, 2014.

\bibitem[Badieirostami et~al.(2010)Badieirostami, Lew, Thompson, and
  Moerner]{badieirostami2010three}
M.~Badieirostami, M.~D. Lew, M.~A. Thompson, and W.~Moerner,
  ``Three-dimensional localization precision of the double-helix point spread
  function versus astigmatism and biplane,'' \emph{Appl. Physics Letters},
  vol.~97, no.~16, p. 161103, 2010.

\bibitem[Nehme et~al.(2020{\natexlab{a}})Nehme, Freedman, Gordon, Ferdman,
  Weiss, Alalouf, Naor, Orange, Michaeli, and Shechtman]{nehme2020deepstorm3d}
E.~Nehme, D.~Freedman, R.~Gordon, B.~Ferdman, L.~E. Weiss, O.~Alalouf, T.~Naor,
  R.~Orange, T.~Michaeli, and Y.~Shechtman, ``Deepstorm3d: dense 3d
  localization microscopy and psf design by deep learning,'' \emph{Nature
  Methods}, vol.~17, no.~7, pp. 734--740, 2020.

\bibitem[Nehme et~al.(2020{\natexlab{b}})Nehme, Ferdman, Weiss, Naor, Freedman,
  Michaeli, and Shechtman]{nehme2020learning}
E.~Nehme, B.~Ferdman, L.~E. Weiss, T.~Naor, D.~Freedman, T.~Michaeli, and
  Y.~Shechtman, ``Learning an optimal psf-pair for ultra-dense 3d localization
  microscopy,'' \emph{arXiv preprint arXiv:2009.14303}, 2020.

\bibitem[Cho et~al.(2011)Cho, Paris, Horn, and Freeman]{cho2011blur}
T.~S. Cho, S.~Paris, B.~K. Horn, and W.~T. Freeman, ``Blur kernel estimation
  using the radon transform,'' in \emph{IEEE Comp. Vision and Pattern
  Recognition (CVPR)}, 2011.

\bibitem[Joshi et~al.(2008)Joshi, Szeliski, and Kriegman]{joshi2008psf}
N.~Joshi, R.~Szeliski, and D.~J. Kriegman, ``Psf estimation using sharp edge
  prediction,'' in \emph{IEEE Comp. Vision and Pattern Recognition (CVPR)},
  2008.

\bibitem[Castro and Ojeda-Casta{\~n}eda(2004)]{castro2004asymmetric}
A.~Castro and J.~Ojeda-Casta{\~n}eda, ``Asymmetric phase masks for extended
  depth of field,'' \emph{Appl. Optics}, vol.~43, no.~17, pp. 3474--3479, 2004.

\bibitem[Yang et~al.(2007)Yang, Liu, and Sun]{yang2007optimized}
Q.~Yang, L.~Liu, and J.~Sun, ``Optimized phase pupil masks for extended depth
  of field,'' \emph{Optics Communications}, vol. 272, no.~1, pp. 56--66, 2007.

\bibitem[Zammit et~al.(2014)Zammit, Harvey, and Carles]{zammit2014extended}
P.~Zammit, A.~R. Harvey, and G.~Carles, ``Extended depth-of-field imaging and
  ranging in a snapshot,'' \emph{Optica}, vol.~1, no.~4, pp. 209--216, 2014.

\bibitem[Boominathan et~al.(2020)Boominathan, Adams, Robinson, and
  Veeraraghavan]{boominathan2020phlatcam}
V.~Boominathan, J.~K. Adams, J.~T. Robinson, and A.~Veeraraghavan, ``{Phlatcam:
  D}esigned phase-mask based thin lensless camera,'' \emph{IEEE Trans. Pattern
  Analysis and Machine Intelligence}, vol.~42, no.~7, pp. 1618--1629, 2020.

\bibitem[Lew et~al.(2011)Lew, Lee, Badieirostami, and
  Moerner]{lew2011corkscrew}
M.~D. Lew, S.~F. Lee, M.~Badieirostami, and W.~Moerner, ``Corkscrew point
  spread function for far-field three-dimensional nanoscale localization of
  pointlike objects,'' \emph{Optics Letters}, vol.~36, no.~2, pp. 202--204,
  2011.

\bibitem[Prasad(2013)]{prasad2013rotating}
S.~Prasad, ``Rotating point spread function via pupil-phase engineering,''
  \emph{Optics Letters}, vol.~38, no.~4, pp. 585--587, 2013.

\bibitem[Wu et~al.(2019)Wu, Boominathan, Chen, Sankaranarayanan, and
  Veeraraghavan]{wu2019phasecam3d}
Y.~Wu, V.~Boominathan, H.~Chen, A.~Sankaranarayanan, and A.~Veeraraghavan,
  ``Phasecam3d—{L}earning phase masks for passive single view depth
  estimation,'' in \emph{IEEE Intl. Conf. Computational Photography (ICCP)},
  2019.

\bibitem[Chang and Wetzstein(2019)]{chang2019deep}
J.~Chang and G.~Wetzstein, ``Deep optics for monocular depth estimation and 3d
  object detection,'' in \emph{IEEE Intl. Conf. Computer Vision (ICCV)}, 2019.

\bibitem[Huang et~al.(2008)Huang, Wang, Bates, and Zhuang]{huang2008three}
B.~Huang, W.~Wang, M.~Bates, and X.~Zhuang, ``Three-dimensional
  super-resolution imaging by stochastic optical reconstruction microscopy,''
  \emph{Science}, vol. 319, no. 5864, pp. 810--813, 2008.

\bibitem[Betzig et~al.(2006)Betzig, Patterson, Sougrat, Lindwasser, Olenych,
  Bonifacino, Davidson, Lippincott-Schwartz, and Hess]{betzig2006imaging}
E.~Betzig, G.~H. Patterson, R.~Sougrat, O.~W. Lindwasser, S.~Olenych, J.~S.
  Bonifacino, M.~W. Davidson, J.~Lippincott-Schwartz, and H.~F. Hess, ``Imaging
  intracellular fluorescent proteins at nanometer resolution,'' \emph{Science},
  vol. 313, no. 5793, pp. 1642--1645, 2006.

\bibitem[Hess et~al.(2006)Hess, Girirajan, and Mason]{hess2006ultra}
S.~T. Hess, T.~P. Girirajan, and M.~D. Mason, ``Ultra-high resolution imaging
  by fluorescence photoactivation localization microscopy,'' \emph{Biophysical
  J.}, vol.~91, no.~11, pp. 4258--4272, 2006.

\bibitem[Juette et~al.(2008)Juette, Gould, Lessard, Mlodzianoski, Nagpure,
  Bennett, Hess, and Bewersdorf]{juette2008three}
M.~F. Juette, T.~J. Gould, M.~D. Lessard, M.~J. Mlodzianoski, B.~S. Nagpure,
  B.~T. Bennett, S.~T. Hess, and J.~Bewersdorf, ``Three-dimensional sub--100 nm
  resolution fluorescence microscopy of thick samples,'' \emph{Nature Methods},
  vol.~5, no.~6, pp. 527--529, 2008.

\bibitem[Girkin and Carvalho(2018)]{girkin2018light}
J.~M. Girkin and M.~T. Carvalho, ``The light-sheet microscopy revolution,''
  \emph{J. Optics}, vol.~20, no.~5, p. 053002, 2018.

\bibitem[Di~Giovanna et~al.(2018)Di~Giovanna, Tibo, Silvestri,
  M{\"u}llenbroich, Costantini, Mascaro, Sacconi, Frasconi, and
  Pavone]{di2018whole}
A.~P. Di~Giovanna, A.~Tibo, L.~Silvestri, M.~C. M{\"u}llenbroich,
  I.~Costantini, A.~L.~A. Mascaro, L.~Sacconi, P.~Frasconi, and F.~S. Pavone,
  ``Whole-brain vasculature reconstruction at the single capillary level,''
  \emph{Scientific Reports}, vol.~8, no.~1, pp. 1--11, 2018.

\bibitem[Lugo-Hernandez et~al.(2017)Lugo-Hernandez, Squire, Hagemann, Brenzel,
  Sardari, Schlechter, Sanchez-Mendoza, Gunzer, Faissner, and
  Hermann]{lugo20173d}
E.~Lugo-Hernandez, A.~Squire, N.~Hagemann, A.~Brenzel, M.~Sardari,
  J.~Schlechter, E.~H. Sanchez-Mendoza, M.~Gunzer, A.~Faissner, and D.~M.
  Hermann, ``3d visualization and quantification of microvessels in the whole
  ischemic mouse brain using solvent-based clearing and light sheet
  microscopy,'' \emph{J. Cerebral Blood Flow \& Metabolism}, vol.~37, no.~10,
  pp. 3355--3367, 2017.

\bibitem[St.~Croix et~al.(2005)St.~Croix, Shand, and Watkins]{st2005confocal}
C.~M. St.~Croix, S.~H. Shand, and S.~C. Watkins, ``Confocal microscopy:
  comparisons, applications, and problems,'' \emph{Biotechniques}, vol.~39,
  no.~6, pp. S2--S5, 2005.

\bibitem[Kelch et~al.(2015)Kelch, Bogle, Sands, Phillips, LeGrice, and
  Dunbar]{kelch2015organ}
I.~D. Kelch, G.~Bogle, G.~B. Sands, A.~R. Phillips, I.~J. LeGrice, and P.~R.
  Dunbar, ``Organ-wide 3d-imaging and topological analysis of the continuous
  microvascular network in a murine lymph node,'' \emph{Scientific Reports},
  vol.~5, no.~1, pp. 1--19, 2015.

\bibitem[Bassi et~al.(2011)Bassi, Fieramonti, D'Andrea, Valentini, and
  Mione]{bassi2011vivo}
A.~Bassi, L.~Fieramonti, C.~D'Andrea, G.~Valentini, and M.~Mione, ``In vivo
  label-free three-dimensional imaging of zebrafish vasculature with optical
  projection tomography,'' \emph{J. Biomedical Optics}, vol.~16, no.~10, p.
  100502, 2011.

\bibitem[Makita et~al.(2006)Makita, Hong, Yamanari, Yatagai, and
  Yasuno]{Makita:06}
S.~Makita, Y.~Hong, M.~Yamanari, T.~Yatagai, and Y.~Yasuno, ``Optical coherence
  angiography,'' \emph{Opt. Express}, vol.~14, no.~17, pp. 7821--7840, 2006.

\bibitem[Grover et~al.(2011)Grover, Quirin, Fiedler, and
  Piestun]{grover2011photon}
G.~Grover, S.~Quirin, C.~Fiedler, and R.~Piestun, ``Photon efficient
  double-helix psf microscopy with application to 3d photo-activation
  localization imaging,'' \emph{Biomedical Optics Express}, vol.~2, no.~11, pp.
  3010--3020, 2011.

\bibitem[Gustavsson et~al.(2018)Gustavsson, Petrov, Lee, Shechtman, and
  Moerner]{gustavsson20183d}
A.-K. Gustavsson, P.~N. Petrov, M.~Y. Lee, Y.~Shechtman, and W.~Moerner, ``3d
  single-molecule super-resolution microscopy with a tilted light sheet,''
  \emph{Nature Communications}, vol.~9, no.~1, pp. 1--8, 2018.

\bibitem[Bennett et~al.(2020)Bennett, Gustavsson, Bayas, Petrov, Mooney,
  Moerner, and Jackson]{bennett2020novel}
H.~W. Bennett, A.-K. Gustavsson, C.~A. Bayas, P.~N. Petrov, N.~Mooney,
  W.~Moerner, and P.~K. Jackson, ``Novel fibrillar structure in the inversin
  compartment of primary cilia revealed by 3d single-molecule superresolution
  microscopy,'' \emph{Molecular Biology of the Cell}, vol.~31, no.~7, pp.
  619--639, 2020.

\bibitem[Quirin and Piestun(2013)]{quirin2013depth}
S.~Quirin and R.~Piestun, ``Depth estimation and image recovery using
  broadband, incoherent illumination with engineered point spread functions,''
  \emph{Appl. Optics}, vol.~52, no.~1, pp. A367--A376, 2013.

\bibitem[Berlich et~al.(2016)Berlich, Br{\"a}uer, and
  Stallinga]{berlich2016single}
R.~Berlich, A.~Br{\"a}uer, and S.~Stallinga, ``Single shot three-dimensional
  imaging using an engineered point spread function,'' \emph{Optics Express},
  vol.~24, no.~6, pp. 5946--5960, 2016.

\bibitem[Wang et~al.(2017)Wang, Cai, Liang, Zhou, Yan, Dan, Bianco, Lei, and
  Yao]{wang2017single}
Z.~Wang, Y.~Cai, Y.~Liang, X.~Zhou, S.~Yan, D.~Dan, P.~R. Bianco, M.~Lei, and
  B.~Yao, ``Single shot, three-dimensional fluorescence microscopy with a
  spatially rotating point spread function,'' \emph{Biomedical Optics Express},
  vol.~8, no.~12, pp. 5493--5506, 2017.

\bibitem[Roider et~al.(2014)Roider, Jesacher, Bernet, and
  Ritsch-Marte]{roider2014axial}
C.~Roider, A.~Jesacher, S.~Bernet, and M.~Ritsch-Marte, ``Axial
  super-localisation using rotating point spread functions shaped by
  polarisation-dependent phase modulation,'' \emph{Optics Express}, vol.~22,
  no.~4, pp. 4029--4037, 2014.

\bibitem[Ikoma et~al.(2021)Ikoma, Kudo, Peng, Broxton, and
  Wetzstein]{ikoma2021deep}
H.~Ikoma, T.~Kudo, Y.~Peng, M.~Broxton, and G.~Wetzstein, ``Deep learning
  multi-shot 3d localization microscopy using hybrid optical--electronic
  computing,'' \emph{Optics Letters}, vol.~46, no.~24, pp. 6023--6026, 2021.

\bibitem[Piestun et~al.(2000)Piestun, Schechner, and
  Shamir]{piestun2000propagation}
R.~Piestun, Y.~Y. Schechner, and J.~Shamir, ``Propagation-invariant wave fields
  with finite energy,'' \emph{JOSA A}, vol.~17, no.~2, pp. 294--303, 2000.

\bibitem[Yanny et~al.(2020)Yanny, Antipa, Liberti, Dehaeck, Monakhova, Liu,
  Shen, Ng, and Waller]{yanny2020miniscope3d}
K.~Yanny, N.~Antipa, W.~Liberti, S.~Dehaeck, K.~Monakhova, F.~L. Liu, K.~Shen,
  R.~Ng, and L.~Waller, ``Miniscope3d: optimized single-shot miniature 3d
  fluorescence microscopy,'' \emph{Light: Science \& Applications}, vol.~9,
  no.~1, pp. 1--13, 2020.

\bibitem[Xue et~al.(2020)Xue, Davison, Boas, and Tian]{xue2020single}
Y.~Xue, I.~G. Davison, D.~A. Boas, and L.~Tian, ``Single-shot 3d wide-field
  fluorescence imaging with a computational miniature mesoscope,''
  \emph{Science Advances}, vol.~6, no.~43, p. eabb7508, 2020.

\bibitem[Zhang et~al.(2020)Zhang, Kellman, Bostan, and Waller]{zhang20203d}
K.~Zhang, M.~R. Kellman, E.~Bostan, and L.~Waller, ``3d fluorescence
  deconvolution with deep priors (conference presentation),'' in
  \emph{Three-Dimensional and Multidimensional Microscopy: Image Acquisition
  and Processing XXVII}, 2020.

\bibitem[Kingma and Ba(2014)]{kingma2014adam}
D.~P. Kingma and J.~Ba, ``Adam: A method for stochastic optimization,''
  \emph{arXiv preprint arXiv:1412.6980}, 2014.

\bibitem[Hamarneh and Jassi(2010)]{hamarneh2010vascusynth}
G.~Hamarneh and P.~Jassi, ``Vascusynth: Simulating vascular trees for
  generating volumetric image data with ground-truth segmentation and tree
  analysis,'' \emph{Computerized Medical Imaging and Graphics}, vol.~34, no.~8,
  pp. 605--616, 2010.

\bibitem[Jassi and Hamarneh(2011)]{jassi2011vascusynth}
P.~Jassi and G.~Hamarneh, ``Vascusynth: Vascular tree synthesis software,''
  \emph{Insight J.}, 2011.

\bibitem[Wang et~al.(2003)Wang, Simoncelli, and Bovik]{wang2003multiscale}
Z.~Wang, E.~P. Simoncelli, and A.~C. Bovik, ``Multiscale structural similarity
  for image quality assessment,'' in \emph{Asilomar Conf. Signals, Systems \&
  Computers}, 2003.

\bibitem[Peruz(2022)]{peruz}
S.~Peruz, ``Shape from focus,''
  \url{https://www.mathworks.com/matlabcentral/fileexchange/55103-shape-from-focus},
  2022.

\bibitem[Goodman(2005)]{goodman2005introduction}
J.~W. Goodman, ``Introduction to fourier optics, roberts \& co,''
  \emph{Publishers, Englewood, Colorado}, 2005.

\bibitem[Bando et~al.(2008)Bando, Chen, and Nishita]{bando2008extracting}
Y.~Bando, B.-Y. Chen, and T.~Nishita, ``Extracting depth and matte using a
  color-filtered aperture,'' in \emph{ACM SIGGRAPH Asia 2008 papers}, 2008, pp.
  1--9.

\bibitem[nan()]{nanoscribe}
\BIBentryALTinterwordspacing
Nanoscribe gmbh. [Online]. Available: \url{https://www.nanoscribe.de/}
\BIBentrySTDinterwordspacing

\end{thebibliography}













\section{Supplementary: Methods}
\subsection{Properties of GL-based rotating PSFs}
\cite{piestun2000propagation} presented a theory on paraxial rotating beams using GL modes, along with a detailed estimation of rotation range, rotation rates, and beam scaling rates. These theoretical values could be extended to give some theoretical description about GL-based rotating PSFs. However, to the best of our knowledge, this has not been shown. Here we present a calculation for the rotation range and rates for GL-based rotating PSFs. 

Consider a monochromatic paraxial beam with width $w_0$ and wavelength $\lambda$, which is generated from GL modes that lie on a single line in the GL modal plane with a slope $V_1$, expressed as:
\begin{equation}
    n_j = V_1m_j + V_2 \ \ \ \ j=1,2,...
\end{equation}
where $V_1, V_2$ are integer constants, and $\{m_j\}$ are non-negative numbers in arithmetic progression. \cite{piestun2000propagation} showed that such a paraxial beam rotates at a rate 
\begin{equation}
    \frac{d\phi}{dz} = \frac{V_1}{1 + (z/z_R)^2}
\end{equation}
where $z_R = \frac{\pi w_0^2}{\lambda}$ is the Rayleigh length. A point with coordinates $(\rho_0,\phi_0)$ at $z=0$, follows the trajectory \cite{piestun2000propagation}:
\begin{align}
    \rho &= \rho_0 \sqrt{1 + (z/z_R)^2} \\
    \phi &= \phi_0 + V_1(arctan(z/z_R))
\end{align}
which implies that the maximum rotation possible (in one direction) is $V_1(\pi/2)$. 

Thus, for the rotating beam case, over a distance of $z_R$, the beam rotates $V_1(\pi/4)$. The way to find out rotation and scaling rates for a GL-based rotating PSF is to find the equivalent Rayleigh length $z_R$ for it in a $4f$ optical system. Note that the GL modes arise from modal solutions of the Fresnel diffraction operator\cite{piestun2000propagation}. The Fresnel diffraction integral is given as follows \cite{goodman2005introduction}:
\begin{equation} \label{eq:fresnel}
    U(u,v) = \frac{e^{jkz}}{j\lambda z}e^{j\frac{k}{2z}(u^2+v^2)}\mathcal{F}\{U(x,y)e^{j\frac{k}{2z}(x^2+y^2)}\}_{\frac{u}{\lambda z},\frac{v}{\lambda z}}
\end{equation}
where $j^2=-1$ and $k$ is the wave number. Eqn \ref{eq:fresnel} determines the wave profile $U(u,v)$ at $z$ distance away from an input wave profile $U(x,y)$. Thus, the phase term that determines the \emph{defocus} or \emph{propagation} by a distance $z$ is 
\begin{equation}\label{eq:rl-beam}
\frac{k}{2z}(u^2+v^2)
\end{equation}
Similarly, upon solving the propagation integrals for a $4f$ system with defocus, the phase term that affects the defocus is given by 
\begin{equation}\label{eq:rl-4f}
    \text{exp}\left(j\frac{k}{2f}(\frac{\Delta z}{f})(u^2+v^2)\right)
\end{equation}
where $\Delta z$ is the distance from the in-focus plane. 
Comparing Eqns \ref{eq:rl-beam} and \ref{eq:rl-4f}, we can obtain the equivalent \textit{Rayleigh length} in the $4f$ system (say $z'_R$) 
\begin{align}
    \frac{1}{z_R} = \frac{z'_R}{f^2}
    \implies z'_R = f^2/z_R = \frac{\lambda f^2}{\pi w_0^2}
\end{align}
Thus, given a GL-based rotating PSF mask designed with beam width $w_0$, for wavelength $\lambda$ and $V_1$ slope in the GL modal plane:
\begin{itemize}
    \item The total rotation amount in one direction of defocus is $V_1(\pi/2)$
    \item The angle of the rotating PSF $\phi(z) = \phi_0 +  V_1(\text{arctan}(\frac{z}{\lambda f^2/\pi w_0^2}))$
    \item And over a depth of $\frac{\lambda f^2}{\pi w_0^2}$ the PSF will rotate by $V_1(\pi/4)$ radians. 
\end{itemize}

\subsection{Analysis of CRLB$_{\phi}$}
\begin{figure}[h!]
    \centering
    \includegraphics{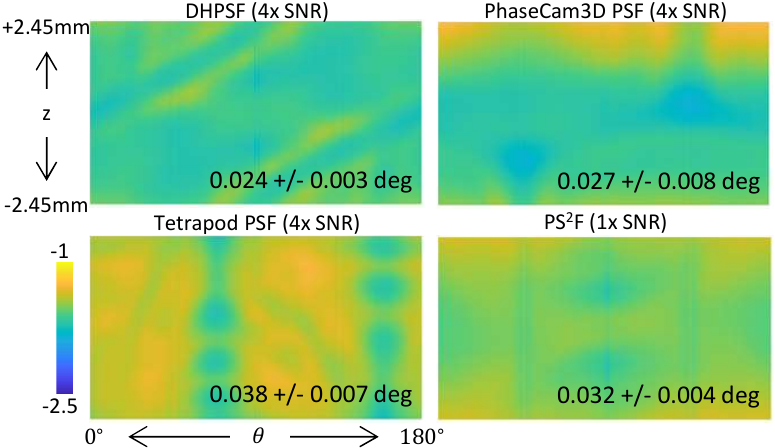}
    \caption{\textbf{CRLB$_{\phi}$ comparison for various phase masks.} The figure shows $\sqrt{CRLB_{\phi}}$ log-scale plots for line images, as a function of line depth $(z)$ and line orientation $(\theta)$. Individual insets depict the mean and standard deviation of the $\sqrt{\textrm{CRLB}_{\phi}}$ values, which are all very low.}
    \label{fig:crlb-phi}
\end{figure}
\begin{figure*}[!t]
\centering
\includegraphics[width=\textwidth]{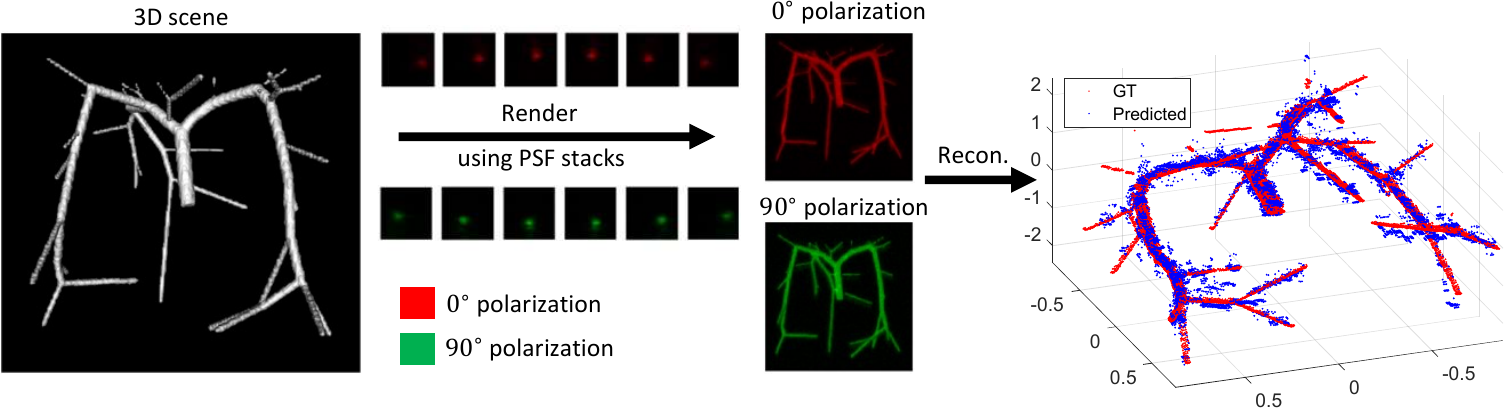}
\caption{\textbf{Simulation methodology to evaluate PS$^2$F}. Given a 3D volume, we used a simulated stack of depth-dependent PSFs to obtain the two polarization images. We then computed the depth at each spatial point with the two polarization images as inputs, resulting in the point cloud shown on the right.}
\label{fig:sims-vasc}
\end{figure*}

\begin{table*}[!t]
\setlength{\tabcolsep}{.25cm}
\renewcommand{\arraystretch}{1.3}
\caption{Depth Estimation performance, averaged over 10 scenes for each scene complexity (noise level$=0.02$). $B$ refers to the number of bifurcations (B) in the 3D vascular structure. Showing mean absolute error (MAE) and RMS error (RMSE) values (all units in mm), and multi-scale SSIM~\cite{wang2003multiscale} (MS-SSIM) values. For the first two metrics, lower is better (indicated by $\downarrow$), and for MS-SSIM, higher is better (indicated by $\uparrow$).}
\centering
\begin{tabular}{|c||c|c|c|c|c|c|c|c|c|}
\hline
 & \multicolumn{3}{c|}{PS$^2$F(proposed)} & \multicolumn{3}{c|}{DHPSF(x2)~\cite{pavani2009three}} & \multicolumn{3}{c|}{TetrapodPSF(x2)~\cite{shechtman2014optimal}}\\
\hline\hline
 & MAE $\downarrow$ & RMSE $\downarrow$ & MS-SSIM $\uparrow$ & MAE $\downarrow$ & RMSE $\downarrow$ & MS-SSIM $\uparrow$ & MAE $\downarrow$ & RMSE $\downarrow$ & MS-SSIM $\uparrow$\\
 \hline
B=6 & \textbf{0.198} & \textbf{0.292} & \textbf{0.990} & 0.284 & 0.46 & \textbf{0.990} & 0.352 & 0.541 & \textbf{0.990}\\
\hline
B=16 & \textbf{0.225} & \textbf{0.334} & \textbf{0.980} & 0.494 & 0.781 & 0.979 & 0.532 & 0.811 &  0.979\\
\hline
B=26 & \textbf{0.211} & \textbf{0.321} & \textbf{0.972} & 0.507 & 0.839 & 0.971 & 0.574 & 0.872 & 0.970\\
\hline
B=36 & \textbf{0.224} & \textbf{0.344} & \textbf{0.969} & 0.548 & 0.904 & 0.967 & 0.585 & 0.906 & 0.966\\
\hline
B=46 & \textbf{0.261} & \textbf{0.413} & \textbf{0.966} & 0.564 & 0.933 & 0.964 & 0.769 & 1.129 & 0.963\\
\hline
\end{tabular}
\label{tab:sim-compare}
\end{table*}
In the CRLB analysis performed in Sec 3.3, we consider an analysis for CRLB$_z$ only. Intuitively, the estimation of line orientation from the image of a line patch is an easier task as compared to estimating line depth. This is because line orientation can be readily estimated from the global structure present in the line image. To verify this, we perform CRLB$_{\phi}$ calculations with the same parameters ($N=100,000$ photon, $\beta=5$ photons/pixel). The $\sqrt{\textrm{CRLB}_{\phi}}$ plots are shown in Supplementary Fig.~\ref{fig:crlb-phi}. The $\sqrt{\textrm{CRLB}_{\phi}}$ values are very low, indicating that our intuition about line orientation estimation being an easier problem is correct.


\section{Supplementary: Simulations}
\subsection{VascuSynth dataset simulations}
The VascuSynth dataset simulations pipeline is best illustrated by Fig.~\ref{fig:sims-vasc}. 

Table~\ref{tab:sim-compare} shows the comparative results for the proposed PS$^2$F, DHPSF, and the Tetrapod PSF over three metrics - Mean Absolute Depth Error, RMS Depth Error, and multi-scale SSIM~\cite{wang2003multiscale}.

\begin{figure}[h!]
\centering
\includegraphics[width=0.8\columnwidth]{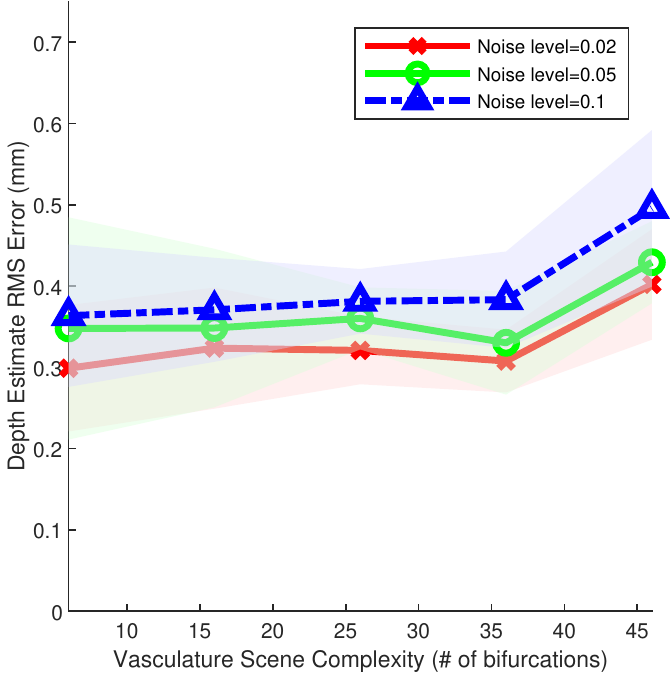}
\caption{Depth Estimation performance, for different noise levels (averaged over 5 scenes for each complexity level, shaded regions show variance across the 5 scenes). This shows the robustness of the proposed PSF and technique.}
\label{fig:sim-noise-analysis}
\end{figure}

We also perform simulations under varying levels of noise, to judge the robustness of the proposed PS$^2$F. We simulate vasculature scene renderings with Poisson noise, and also Gaussian read-out noise of levels $0.02$, $0.05$, and $0.1$ - which correspond to PSNRs of 34dB, 26dB, and 20dB. Fig. \ref{fig:sim-noise-analysis} shows that with increasing noise, the proposed PS$^2$F reconstruction only worsens slightly, showing the robustness of the proposed PSF and method to noise.

\subsection{Ablation study: Comparison with polarized stereo pair}
\begin{figure}[t!]
    \centering
    \includegraphics[width=\columnwidth]{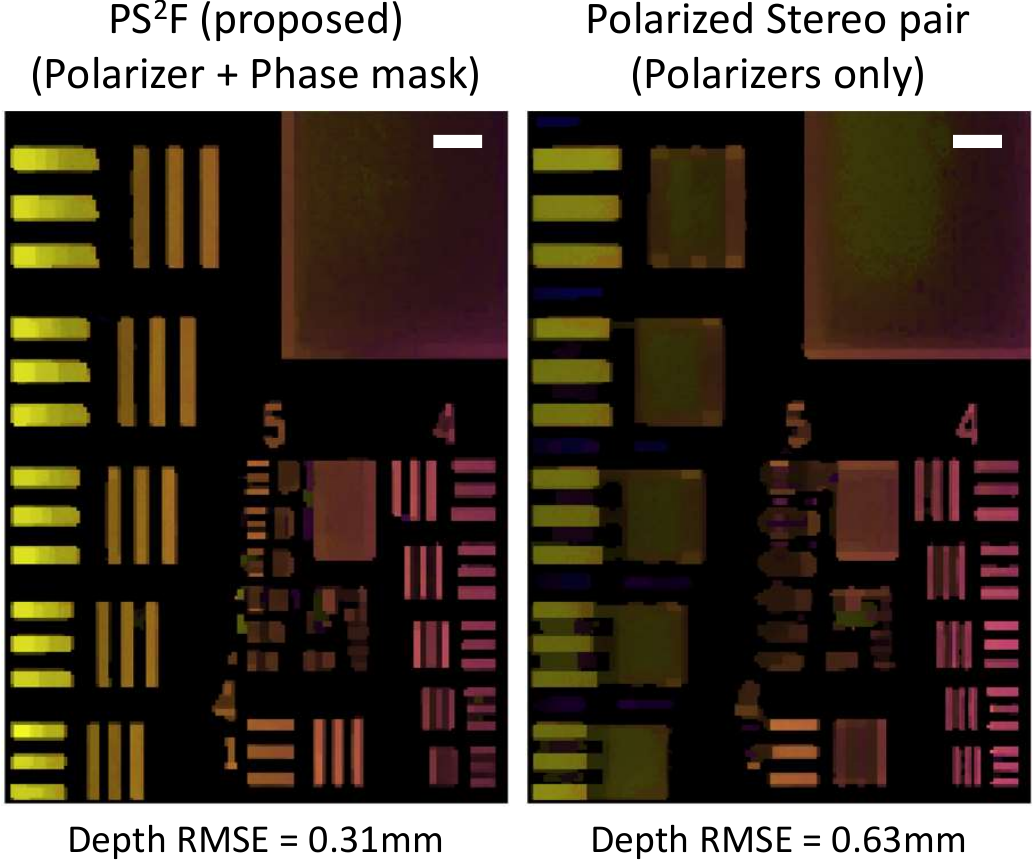}
    \caption{\textbf{Comparing polarizer-phase mask design vs. polarizer-only design:} Simulated 3D USAF target reconstructions shown for the PS$^2$F case and for the polarized-stereo (polarizers-only) case. Showing depth color-coded reconstructed intensity plots. A polarizer-phase mask design gives a $\sim2\times$ lower depth RMS error in reconstruction as compared to the polarizer-only design, and poorer spatial resolution. Scale bars indicate $0.1$~mm.}
    \label{fig:ablation}
\end{figure}

The proposed PS$^2$F has been created using a novel polarizer-phase mask design and the usage of a polarization-based camera sensor. We perform an ablation study, comparing our proposed PS$^2$F (polarizer-phase mask design) with a PSF created using just polarizer halves in the Fourier (pupil) plane. This creates images in the $s$, $p$-polarized channels that correspond to different half apertures (taking inspiration from \cite{bando2008extracting}). With such a polarized stereo-vision effect, where each polarization channel sees a slightly different perspective, we attempt reconstruction of a 3D USAF skew target in simulation. The reconstructed intensity plots (depth color-coded) for the PS$^2$F and the Polarized-Stereo PSF are shown in Fig.~\ref{fig:ablation}. A polarizer-phase mask design gives better spatial resolution, and a $\sim2\times$ lower depth RMS error in reconstruction as compared to the polarizer-only design. This can possibly be attributed to the fact that the synthetic polarized-stereo PSF pair has a narrow baseline length, and also that the PSFs are no longer compact, leading to worse spatial resolution.   

\section{Supplementary: Experiments}
\subsection{Pupil plane encoding using polarizers and phase mask}
\begin{figure}[!t]
\centering
\includegraphics[width=\columnwidth]{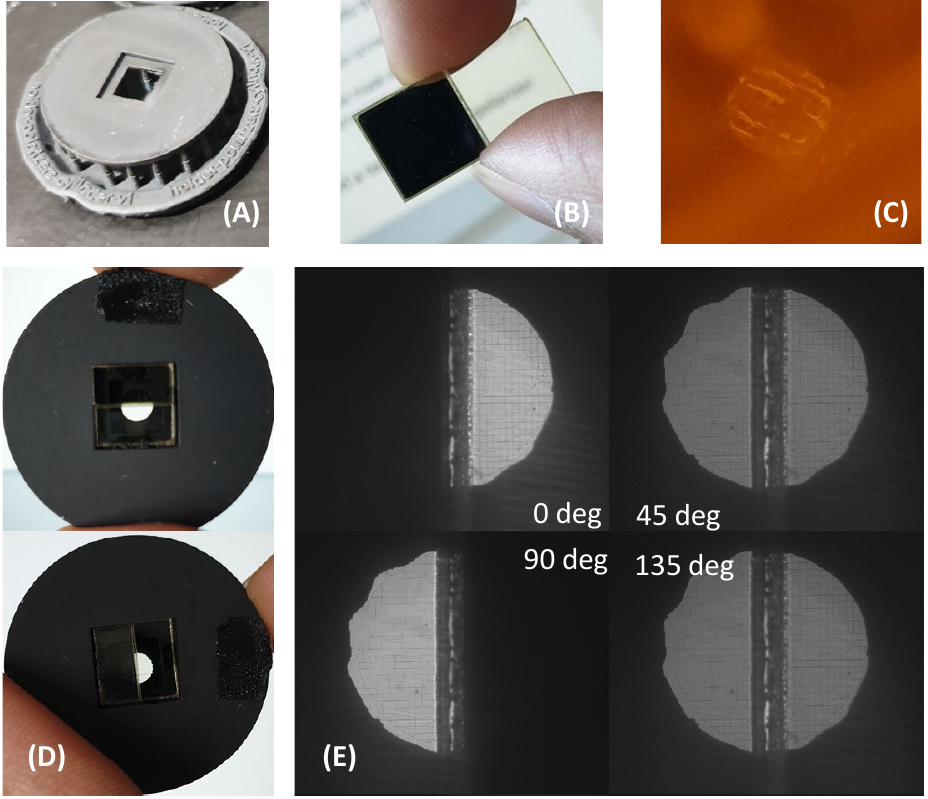}
\caption{\textbf{Polarizer-mask setup in the pupil plane.} (A) shows the 3D printed holder, in which two polarizer films (B) aligned to orthogonal polarizations (0 and 90 degree) will be placed, along with the fabricated DHPSF mask (C). (D) shows the entire setup kept in front of an LCD screen with polarized illumination. This demonstrates how such a setup allows for light from only one half of the mask to pass through. (E) shows an image of the polarizer-phase mask after keeping it in the pupil plane (imaged using a polarization camera).}
\label{fig:polarizer-mask-setup}
\end{figure}
Figure~\ref{fig:polarizer-mask-setup} illustrates how the polarizer-phase mask encoding was achieved. 

\subsection{Phase mask fabrication}
We used a two-photon photolithography 3D printer (Photonic Professional \textit{GT}, Nanoscribe GmbH~\cite{nanoscribe}) to print the phase mask structures using IP-Dip photoresist on a $700~\mu$m thick fused silica substrate. A $3$~mm diameter phase mask corresponding to the DHPSF phase mask profile was fabricated, with $2~\mu$m discretization in the $xy$-dimensions. For reliable fabrication, the phase profile was also quantized to 5 different levels. 

\subsection{Fluorescent sample preparation}
The fluorescent bead sample was prepared by adding 10$\mu$m fluorescent beads solutions with varying concentrations in polydimethylsiloxane (PDMS) (Sylgard, Dow Corning; 10:1 elastomer:cross-linker weight ratio). The fluorescent cotton strand 3D sample was prepared by using a fluorescent highlighter on a tiny piece of a cotton ball, which consisted of strands with thicknesses ranging from $10$-$30~\mu$m. The fluorescent-highlighted sample was then placed in a PDMS solution. All samples were cured at room temperature for a minimum of 24 hours.\\

\subsection{Hyperparameter selection}
In reconstruction, the optimization objective (Eq 5 in the main paper) has two hyperparameters - $\lambda_{TV}$ and $\lambda_{L1}$ - these control total variation (TV) and L1 regularization. The three broad types of experimental reconstructions shown in the submission (USAF target, fluorescent beads, fluorescent cotton strands) had different scene settings and properties, such as transmission vs. fluorescence illumination (different background signal statistics), extended vs. point-like scenes (different sparsity) and thus the hyperparameters were set accordingly to -
\begin{itemize}
    \item USAF (planar) target: $(\lambda_{L1}, \lambda_{TV})=(0.02, 0.005)$ 
    \item Fluorescent beads: $(\lambda_{L1}, \lambda_{TV})=(0.05, 0.00)$
    \item Fluorescent strands: $(\lambda_{L1}, \lambda_{TV})=(0.002, 0.002)$
\end{itemize}
For the same type of scene, high sensitivity to hyperparameter weights was not observed in our reconstructions.



\ifpeerreview \else

\fi

\end{document}